\documentclass[prb,twocolumn,aps,floatfix,groupedaddress,longbibliography,nofootinbib]{revtex4-2}
\usepackage{stix}
\usepackage{amsmath}
\usepackage{siunitx}
\usepackage{graphicx}
\usepackage[usenames,dvipsnames]{xcolor}
\usepackage{bm}
\usepackage{hyperref}
\usepackage{float}
\usepackage{braket}
\usepackage[utf8]{inputenc}
\usepackage{lipsum}     
\usepackage[most]{tcolorbox}
\usepackage{blindtext}
\hypersetup{
  colorlinks=true,
  citecolor=blue,
  linkcolor=blue,
  urlcolor=blue}

\newcommand{\refeq}[1]{Eq.~(\ref{#1})}

\newcommand{\refeqand}[2]{Eqs.~(\ref{#1}) and (\ref{#2})}
\newcommand{\reffig}[1]{Fig.~\ref{#1}}

\newcommand{\refsec}[1]{Sec.~\ref{#1}}
\newcommand{\refapp}[1]{Appendix~\ref{#1}}

\newcommand{\punc}[1]{\,{\text{#1}}}
\newcommand{\sub}[1]{_{\text{#1}}}

\newcommand{\super}[1]{^{\text{#1}}}

\newcommand{\sprC}{^{\{\mathrm{C}\}}}

\newcommand{\DTO}{Dy$_2$Ti$_2$O$_7$}

\begin{document}
	

    \title{Crossover from string to cluster dynamics following a field quench in spin ice}
       
	\author{Sukla Pal}
    \affiliation{Department of Physics, Centre for Quantum Science, and The Dodd-Walls Centre for \\ Photonic and Quantum Technologies, University of Otago, Dunedin 9016, New Zealand}
    \affiliation{School of Physics and Astronomy, The University of Nottingham, Nottingham, NG7 2RD, United Kingdom}

	\author{Stephen Powell}
	\affiliation{School of Physics and Astronomy, The University of Nottingham, Nottingham, NG7 2RD, United Kingdom}

	\date{\today}
	
\begin{abstract}
	We investigate quench dynamics of spin ice after removal of a strong magnetic field along the \([100]\) crystal direction, using Monte Carlo simulations and theoretical arguments. We show how the early-time relaxation of the magnetization can be understood in terms of nucleation and growth of strings of flipped spins, in agreement with an effective stochastic model that we introduce and solve analytically. We demonstrate a crossover at longer times to a regime dominated by approximately isotropic clusters, which we characterize in terms of their morphology, and present evidence for a percolation transition as a function of magnetization.
\end{abstract}
  	 \maketitle
	\section{Introduction} 

Studying the nonequilibrium dynamics of many-body systems provides a way to explore phenomena that are not accessible at or near equilibrium. The simplest protocol, at least conceptually, is the ``quench'' \cite{Polkovnikov2011,Mitra2018,Castelnovo2010}, where a system is initially prepared in equilibrium and then a sudden change is made to one or more external parameters, such as temperature \cite{Castelnovo2010} or an applied field \cite{Mostame2014b}. Quenches have been realized in numerous experiments, principally in cold atomic gases \cite{Greiner2002,Chen2011, Matthias2011, Nicklas2015, Prufer2018}, but also in magnetic systems \cite{Paulsen2014}.

\begin{figure}[H]
\includegraphics{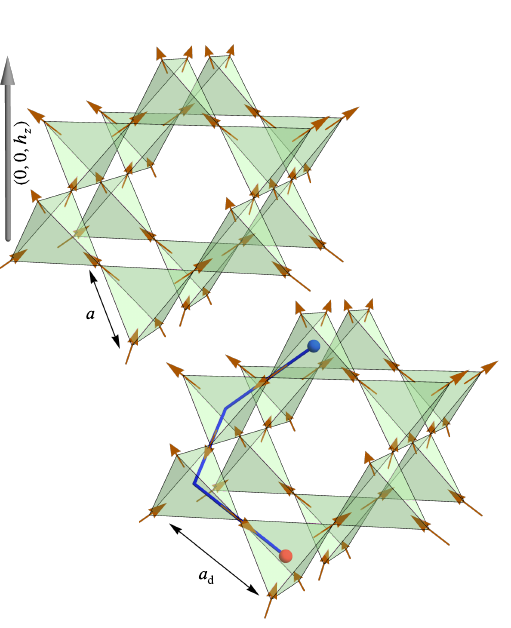}
		\caption{Part of the pyrochlore lattice, a network of corner-sharing tetrahedra, with the \([100]\) crystal direction shown vertically. Top-left: Initial configuration, where all spins are aligned with the magnetic field \(\boldsymbol{h} = (0,0,h_z)\) (gray vertical arrow), as far as possible given the local easy-axis constraints. Bottom-right: Example configuration following the quench of the field to zero. The flipped spins form a ``string'' of length \(\ell = 3\), terminated by monopoles (red and blue spheres) at either end. The tetrahedron centers lie on a diamond lattice with nearest-neighbor distance \(a\sub{d} = \sqrt{3/2}a\),  where \(a\) is the nearest-neighbor distance in the pyrochlore lattice.}
		\label{FigPyrochlore}
	\end{figure}

These include the spin ice materials \cite{Harris1997,Gardiner2010}, a class of frustrated pyrochlore oxides with unusual low-temperature properties. The frustration results from a combination of the pyrochlore lattice (a network of corner-sharing tetrahedra; see \reffig{FigPyrochlore}), strong easy-axis anisotropy, and effectively ferromagnetic nearest-neighbor interactions \cite{Isakov2005,Castelnovo2008} (see \refsec{Sc:Hamiltonian} for details).
  
In the ``classical'' spin ices, such as Dy$_2$Ti$_2$O$_7$ \cite{Ramirez1999} and Ho$_2$Ti$_2$O$_7$ \cite{Rosenkranz2000, Bramwell2001b}, quantum effects are negligible \cite{Rau2015} and the magnetic moments are well approximated as classical Ising degrees of freedom. (By contrast, in ``quantum spin ices'' \cite{Gingras2014} such as Pr$_2$Sn$_2$O$_7$ and Pr$_2$Zr$_2$O$_7$ significant tunneling occurs and a classical description is insufficient.) The spin configurations with lowest energy are those that obey the ``ice rule'' on each tetrahedron: two spins point outwards and two point inwards. Such states form an extensive low-energy manifold, resulting in a large residual entropy at least down to temperatures \(T\simeq \qty{0.35}{\kelvin}\) \cite{Giblin2018}. In this regime, the system behaves as a strongly correlated paramagnet, referred to as a ``Coulomb phase'' \cite{Fennell2009,Henley2010}, while magnetic ordering is predicted to occur at still lower \(T \simeq \qty{0.15}{\kelvin}\) \cite{denHertog2000}.

The elementary excitations above the low-energy manifold are tetrahedra at which the ice rule is broken, where three spins point out and one in, or {\it vice versa}. (Tetrahedra where all four spins point out or in also exist and have still higher energy.) Such excitations, which occur at finite density for any nonzero \(T\), are points where the local magnetization has nonzero divergence and are hence monopoles of the magnetic field \(\boldsymbol{H}\) \cite{Castelnovo2008}. These monopoles are deconfined \cite{Castelnovo2012}, interacting through a magnetic Coulomb law, and can be manipulated by applied magnetic fields \cite{Kadowaki2009,Giblin2011,Mostame2014b}.

In this work, we use Monte Carlo simulations to study dynamics in classical spin ice following a quench of a magnetic field applied along the \([100]\) crystal direction. In the protocol we consider, a large field is initially applied, polarizing the spins along this direction \cite{Fennell2005}, and then suddenly removed, leaving the spins to relax in zero field.

The equilibrium properties of spin ice in a \([100]\) field are naturally described in terms of strings of flipped spins terminated by a monopole at either end \cite{Jaubert2008,Morris2009}. At a critical ratio of temperature to field, there is a crossover where such strings proliferate \cite{Jaubert2008}, which becomes a (``Kasteleyn'' \cite{Kasteleyn1963}) phase transition in the limit where the proliferation temperature is much smaller than the energy cost of a monopole \cite{Jaubert2008,Powell2008,Powell2013}.

The quench that we consider here effectively drives the system across this transition, starting on the low-temperature side. We show that the dynamics is initially driven by the nucleation and growth of strings and derive an effective stochastic model for these processes, which gives quantitatively accurate results at early times. At later times, we observe a crossover from string to cluster dynamics, which we characterize in terms of the size and shape of the largest cluster. We also find a percolation transition analogous to the equilibrium Kasteleyn transition, which appears to demonstrate a crossing point with system size.

The paper is organised as follows. We begin in \refsec{Sc:model} with a description of the model of spin ice and its dynamics. Following that, in \refsec{Sc:qdyn}, we consider the relaxation of bulk properties, specifically the density of magnetic monopoles and the magnetization, following the field quench. In \refsec{Sc:SCdynamics}, we show how the relaxation process can be understood in terms of strings and clusters of flipped spins. The cluster percolation transition is described in \refsec{Sc:perco}. Finally, we conclude in \refsec{Sc:conclusions}, including a discussion of relevance to experiments.
	
	\section{Model and dynamics} 
	\label{Sc:model}

 \subsection{Hamiltonian}
 \label{Sc:Hamiltonian}
	
	We study a model of spin ice with classical spins \(\boldsymbol{S}_i\) of magnetic moment \(\mu\simeq\qty{10}{\mu\sub{B}}\) on the sites \(i\) of a pyrochlore lattice. The spins are constrained to \(\boldsymbol{S}_i = \sigma_i\hat{\boldsymbol{n}}_i\), where \(\sigma_i = \pm 1\) is an effective Ising degree of freedom and the fixed unit vector \(\hat{\boldsymbol{n}}_i\) points along the local \(\langle 111 \rangle\) easy axis between the centers of the two tetrahedra to which each site belongs \cite{Bramwell2001}. More precisely, we define \(\eta_\alpha = \pm 1\) for the two orientations of tetrahedra \(\alpha\) in the pyrochlore lattice, and choose \(\hat{\boldsymbol{n}}_i\) to point from the tetrahedron with \(\eta_\alpha = +1\) to the tetrahedron with \(\eta_\alpha = -1\).
	
	The interactions between the spins are well described by the dipolar spin ice Hamiltonian \cite{denHertog2000}
	\begin{equation}
	    H\sub{DSI} = -J\sum_{\langle i j \rangle} \boldsymbol{S}_i\cdot \boldsymbol{S}_j + D \sum_{i > j} V\sub{dd}\left(\boldsymbol{S}_i,\boldsymbol{S}_j,\frac{\boldsymbol{r}_i - \boldsymbol{r}_j}{a}\right)\punc,
	\end{equation}
	where \(a\) is the nearest-neighbor distance in the pyrochlore lattice, \(D = \mu_0 \mu^2/(4\pi a^3)\) is the dipole energy scale, and
	\begin{equation}
	    V\sub{dd}(\boldsymbol{S},\boldsymbol{S}',\boldsymbol{r}) = \frac{\boldsymbol{S}\cdot\boldsymbol{S}'}{\lvert\boldsymbol{r}\rvert^3} - \frac{3(\boldsymbol{S}\cdot\boldsymbol{r})(\boldsymbol{S}'\cdot\boldsymbol{r})}{\lvert\boldsymbol{r}\rvert^5}
	\end{equation}
	is the interaction energy of a pair of magnetic dipoles. (For simplicity, we neglect further-neighbor exchange interactions, which may become significant at lower temperatures \cite{Fennell2004,Yavorskii2008}.)
 
    For \DTO, the coefficients take values \(J = \qty{-3.72}{\kelvin}\) and \(D = \qty{1.41}{\kelvin}\) \cite{denHertog2000} (we set \(k\sub{B} = 1\) throughout).
    Because of the Ising constraint, for every pair of nearest-neighbor sites \(i\) and \(j\), \(\boldsymbol{S}_i\cdot\boldsymbol{S}_j = -\frac{1}{3}\sigma_i\sigma_j\) while \(V\sub{dd}\left(\boldsymbol{S}_i,\boldsymbol{S}_j,\frac{\boldsymbol{r}_i - \boldsymbol{r}_j}{a}\right) = \frac{5}{3}\sigma_i\sigma_j\) \cite{denHertog2000}. Taking into account both terms in \(H\sub{DSI}\), the net interaction between nearest neighbors can therefore be written as \(-3J\sub{eff}\boldsymbol{S}_i\cdot\boldsymbol{S}_j = +J\sub{eff}\sigma_i\sigma_j\) where
	\begin{equation}
	    J\sub{eff} = \frac{1}{3}J + \frac{5}{3}D \simeq +\qty{1.1}{\kelvin}
	\end{equation}
	for \DTO.
	This antiferromagnetic nearest-neighbor interaction (in terms of \(\sigma_i\)) is frustrated and is minimized by the ``ice rules'' states in which two spins point into each tetrahedron and two point out.
	
	Rather than using the full Hamiltonian \(H\sub{DSI}\), we approximate the dipolar interactions using the dumbbell model \cite{Castelnovo2008}, which replaces each magnetic dipole, of moment \(\mu\boldsymbol{S}_i\), by a pair of magnetic charges \(\pm \mu/a\sub{d}\) at the centers of the two tetrahedra to which that spin belongs. These tetrahedron centers form a diamond lattice with nearest-neighbor distance \(a\sub{d} = \sqrt{3/2}a\) (see \reffig{FigPyrochlore}); in the following, we use \(\alpha\) to label both a tetrahedron and the corresponding diamond site. This approximation, which differs from the DSI model by quadrupolar corrections \cite{Isakov2005}, considerably reduces the computational complexity of the problem and provides an accurate approximation except at very low temperatures \cite{Bovo2018}.
	
	Within this model, the Hamiltonian is given (up to an unimportant constant) by \cite{Castelnovo2008}
	\begin{equation}
	    H = \frac{\nu a\sub{d}^2}{4\mu^2}\sum_\alpha Q_\alpha^2 + \frac{\mu_0}{4\pi}\sum_{\alpha>\beta} \frac{Q_\alpha Q_\beta}{\lvert\boldsymbol{r}_{\alpha}-\boldsymbol{r}_\beta\rvert}
	    \punc,
	\end{equation}
	where \(Q_\alpha\) is the total magnetic charge on diamond site \(\alpha\), and the on-site energy \cite{Raban2019}
	\begin{equation}
	    \nu = \frac{2}{3}J + \frac{8}{3}\left[1 + \sqrt{\frac{2}{3}}\right]D
	\end{equation}
	can be fixed by considering the energy change due to a single spin flip \cite{Castelnovo2008}.
	
	The magnetic charge \(Q_\alpha\) is given by summing the contributions from the four dumbbells representing the four spins on tetrahedron \(\alpha\). We write it as \(Q_\alpha = 2n_\alpha\mu/a\sub{d}\), where
	\begin{equation}
	    n_\alpha = -\frac{1}{2}\eta_\alpha \sum_{i \in \alpha} \sigma_i\punc.
	\end{equation}
	takes integer values \(0\), \(\pm 1\), and \(\pm 2\). (The sum is over sites \(i\) belonging to tetrahedron \(\alpha\).) The Hamiltonian can then be rewritten as \cite{Raban2019}
	\begin{equation}
	\label{eq:H}
	    H = \nu\sum_\alpha n_\alpha^2 + \gamma U\sub{C}\sum_{\alpha > \beta} n_\alpha n_\beta V\sub{C}\left(\frac{\boldsymbol{r}_{\alpha}-\boldsymbol{r}_\beta}{a\sub{d}}\right)\punc,
	\end{equation}
	where \(V\sub{C}(\boldsymbol{r}) = \lvert\boldsymbol{r}\rvert^{-1}\) is the (magnetostatic) Coulomb potential and 
	\begin{equation}
	    U\sub{C} = \frac{\mu_0}{4\pi}\left(\frac{2\mu}{a\sub{d}}\right)^2\frac{1}{a\sub{d}} = \frac{8}{3}\sqrt{\frac{2}{3}}D \simeq \qty{3.1}{\kelvin}\punc,
	\end{equation}
	for \DTO. We include the dimensionless parameter \(\gamma\), which in reality takes the value \(\gamma = 1\), in order to vary the strength of the dipolar interactions in our simulations.

In practice, to calculate the long-range interactions between the magnetic monopoles, we modify the Coulomb potential \(V\sub{C}\) by including mirror charges, implemented using Ewald summation \cite{Leeuw1980, Melko2004,ThesisJaubert}.

	Within the dumbbell model, all configurations that obey the ice rules, i.e., that have \(n_\alpha = 0\) on every tetrahedron \(\alpha\), are degenerate ground states, with energy \(H\sub{gs}=0\). The lowest-energy excited states each have a single spin flipped relative to a ground state, or equivalently a pair of charges \(n_\alpha = \pm 1\) on adjacent tetrahedra. Their energy is therefore \(H\sub{min}=2\nu - \gamma U\sub{C}\), with the two terms coming from the on-site and Coulomb terms in \refeq{eq:H}, respectively. Since \(H\sub{min}\) gives the activation energy for dynamics based on single spin flips, we define \(2\Delta = H\sub{min} - H\sub{gs}\), giving
 \begin{equation}
 \label{Eq:Delta}
 \Delta = \nu - \frac{1}{2}\gamma U\sub{C}\punc.
 \end{equation}

Rather than treating \(\nu\), the on-site interaction, and \(\gamma\), the relative strength of the Coulomb interaction, as independent parameters in our simulations, we choose to fix \(\Delta = \qty{2.8}{\kelvin}\), corresponding to \DTO, while allowing \(\nu\) to vary with \(\gamma\) according to \refeq{Eq:Delta}. Our motivation for this is that we expect the Boltzmann weight for a monopole, \(e^{-\Delta/T}\), to set the principal timescale for the dynamics, and so holding this fixed while varying \(\gamma\) allows us to isolate the effects of the long-range interactions. In the dumbbell picture, it effectively means that changing \(\gamma\) tunes the strength of the Coulomb interaction between further-neighbor tetrahedra but not between nearest neighbors.

\begin{figure} \includegraphics[width=1.0\columnwidth]{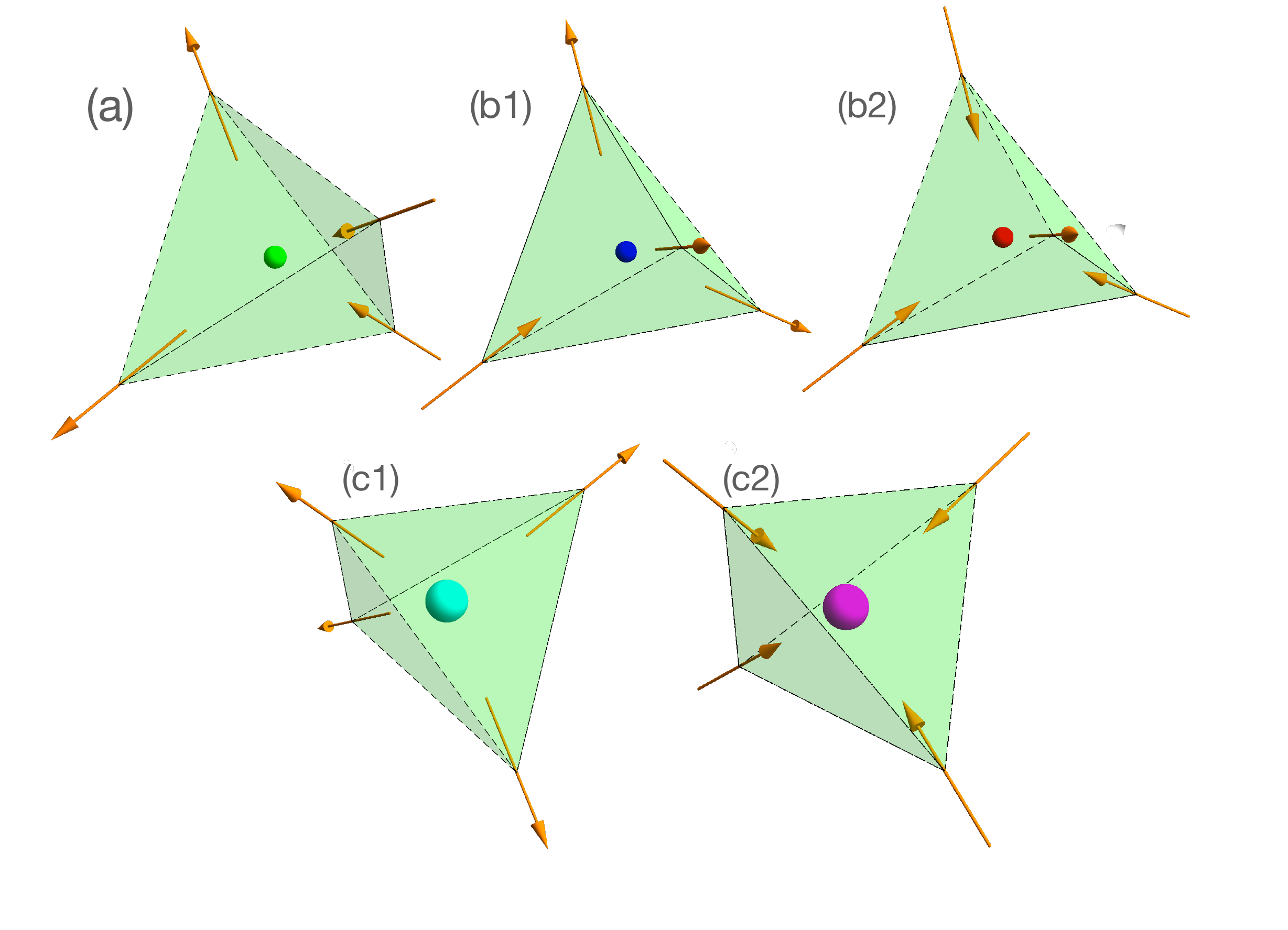}
	\caption{Different classes of tetrahedron configuration, corresponding to vertex types in the diamond lattice. (a) 2 spins in--2 spins out: neutral configuration, with effective magnetic charge \(n_\alpha = 0\) and degeneracy 6. (b1) 3 out--1 in and (b2) 3 in--1 out: single monopole, \(n_\alpha = \pm 1\), with total degeneracy 8 (4 for each sign). (c1) all out and (c2) all in: double monopoles, \(n_\alpha = \pm 2\), with total degeneracy 2.}
 \label{excitation}
\end{figure}

Within the nearest-neighbor model, \(\gamma = 0\), the system can be thought of as a collection of vertices (sites of the diamond lattice) of sixteen  different types. Six of these satisfy the ice rule [neutral, \(n_\alpha = 0\), see \reffig{excitation}(a)], two are all-in or all-out [\(n_\alpha = \pm 2\), see \reffig{excitation}(c1,c2)], and the remaining eight are three-in--one-out or three-out--one-in [\(n_\alpha = \pm 1\), see \reffig{excitation}(b1,b2)]. For \(\gamma > 0\), the configuration energy is no longer simply a sum of vertex terms.

\subsection{Dynamics and simulation parameters}
\label{sec:dynamics}

Our simulations are performed on pyrochlore lattices with \(N\sub{s}=16N\sub{u}^xN\sub{u}^yN\sub{u}^z\) sites, where \(N\sub{u}^{x,y,z}\) are the numbers of cubic unit cells of the fcc Bravais lattice of pyrochlore. Periodic boundary conditions are applied in each direction. The total number of tetrahedra (of both orientations), equal to the number of sites of the diamond lattice, is \(N\sub{d}=\frac{1}{2}N\sub{s}\).

We treat the dynamics using the so-called ``standard model'' \cite{Hallen2022} of uncorrelated spin flips with a single temperature-independent timescale \(\tau\sub{flip}\sim\SI{3}{\milli\second}\) \cite{Ryzhkin2005,Jaubert2009}. Flips are attempted at randomly chosen sites at a rate \(N\sub{s}\tau\sub{flip}^{-1}\), and accepted with a Glauber probability \cite{Glauber1963, Suzen2014, Binder2010}
\begin{equation}
P\sub{G}(\delta E) = \frac{1}{e^{\delta E/T}+1}\punc,
\end{equation}
where \(\delta E\) is the associated change in energy and \(T\) is the temperature. The time \(t\) in our numerical results is therefore effectively measured in units of \(\tau\sub{flip}\).

We consider dynamics following an instantaneous ``quench'' of the applied magnetic field \(\boldsymbol{h}\), which couples to the spins through a Zeeman term \(H\sub{Z} = -\boldsymbol{h}\cdot \boldsymbol{M}\), where
\begin{equation}
\boldsymbol{M} = \sum_i \boldsymbol{S}_i
\end{equation}
is the total magnetization (in units of the ionic magnetic moment \(\mu\)). The initial field is chosen along the \([100]\) crystal direction, which we take as the \(z\) axis, taking the value \(\boldsymbol{h} = (0,0,h_z)\), with magnitude \(h_z \gg T\). As a result, all spins are aligned with the field to the maximum extent consistent with the easy-axis constraint, as illustrated in the top-left panel of Fig.~\ref{FigPyrochlore}, where \([100]\) points upwards. (We assume that \(h_z\) is much smaller than the crystal-field term that enforces the easy-axis constraint, which is of order \(\qty{300}{\kelvin}\) in the classical spin ice materials \cite{Gardiner2010}. For example, an applied field of magnitude \(\mu_0 \lvert \boldsymbol{H}\rvert=\qty{1}{\tesla}\) corresponds to \(\lvert \boldsymbol{h} \rvert \sim \qty{7}{\kelvin}\).) This configuration satisfies the ice rules and so minimizes the (dumbbell-model) Hamiltonian \(H\), since each tetrahedron has its top two spins pointing outward and bottom two spins pointing inward \cite{Fukazawa2002}.

At \(t=0\), the field is instantaneously reduced to \(\boldsymbol{h} = \boldsymbol{0}\), and the subsequent dynamics takes place in zero field starting from the fully magnetized configuration. Since the spin-flip dynamics is ergodic and the (post-quench) Hamiltonian preserves the full symmetry of the lattice, the dynamics progresses towards thermal equilibrium at temperature \(T\), a spin-ice state with a finite density of monopoles and zero mean magnetization.

Our interest in this work is in studying how the thermal equilibrium state is reached and how the nature of the excitations controls the dynamics into the equilibrium state. We first characterize the relaxation to equilibrium by considering bulk properties, before studying how this happens in terms of the appearance of strings and clusters of flipped spins.

\section{Relaxation of bulk properties}
\label{Sc:qdyn}

\subsection{Density of monopoles}

\begin{figure}
\includegraphics[width=\columnwidth]{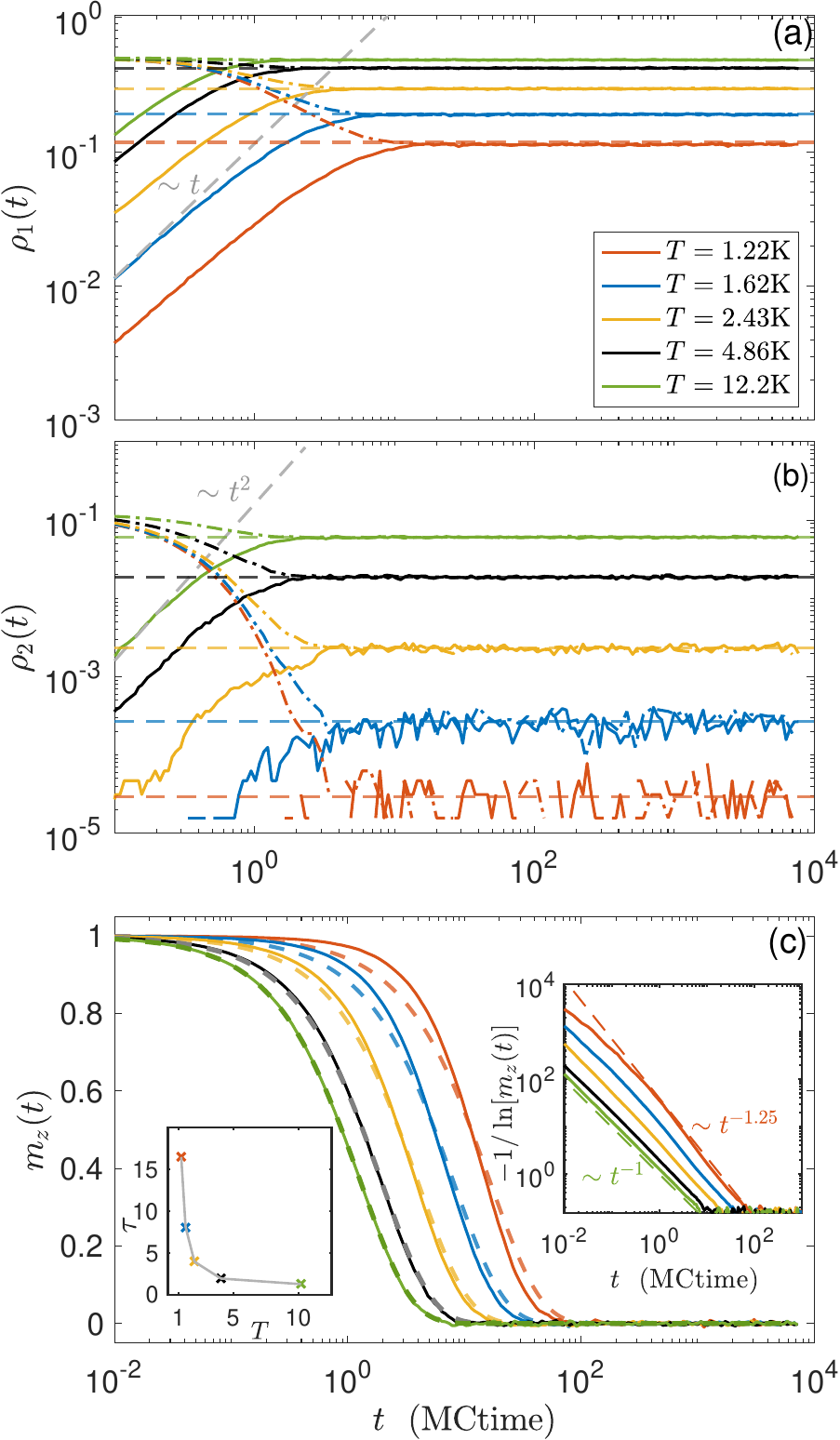}
\caption{Time evolution of density of magnetic monopoles with (a) $|n_{\alpha}| = 1$ (b) $|n_{\alpha}| = 2$ at different temperatures \(T\) for a system of \(N\sub{s}=128\) spins (\(2\times2\times2\) cubic unit cells), averaged over 1000 independent runs. Solid (resp.\ dash--dot) lines corresponds to the fully magnetized (disordered) initial configuration. At long time both densities reach their equilibrium values shown with horizontal dashed lines and obtained from \refeq{Neq}. (c) Time evolution of the \(z\) component of the magnetization, \(m_z(t)\), for the same \(T\) values as in panel (a). Dashed lines show fits to $e^{-t/\tau}$ for each \(T\). The left inset shows a plot of the fitted $\tau$ values, along with a fit to \(\tau \propto \exp(T_0/T)\) with \(T_0 = \qty{3.5}{\kelvin}\). The right inset shows the same data as the main panel with a double-logarithmic vertical axis, on which a stretched exponential \(\sim e^{-(t/\tau)^\beta}\) would appear as a straight line with slope \(-\beta\).}
\label{mdensity}
\end{figure}

The starting configuration, with all spins pointing upward, obeys the ice rule on every tetrahedron, and so has no monopoles. In the long-time limit, the system reaches thermal equilibrium at temperature \(T\) and so monopoles with both \(\lvert n_\alpha \rvert = 1\) and \(2\) will occur at finite density.

Our MC results for the density of monopoles (absolute number of monopoles per tetrahedron) of both types, \(\rho_1\) and \(\rho_2\) respectively, are shown in \reffig{mdensity}(a,b), for nearest-neighbor spin ice, \(\gamma = 0\), with \(N\sub{s}=128\) spins. As expected, both densities increase from zero and quickly reach equilibrium values that increase with temperature.

Starting from an ice-rules configuration, a spin flip produces a pair of monopole excitations of charge \(\pm 1\) on adjacent tetrahedra. We therefore expect \(\rho_1 \sim t\) at very early time, which agrees with our results for \(t \lesssim 0.1\). A charge-\(\pm 2\) monopole can be created by a spin flip at a tetrahedron already containing a monopole. Their density should therefore increase as \(\rho_2 \sim t^2\), which is also consistent with the MC results at similar times.

At long times, the equilibrium densities are determined by the activation energy \(\Delta\), as well as the further neighbor interactions \(\gamma\). For \(\gamma = 0\), the equilibrium densities can be calculated by treating the tetrahedra as independent and considering the Boltzmann weight and multiplicity of each configuration \cite{Levis2013}
	\begin{equation}
    \label{Neq}
	\begin{aligned}
	\rho_{1}\super{eq} &= \frac{8w_{1}}{6 + 8w_1 + 2w_2}\punc,\\
	\rho_2\super{eq} &= \frac{2w_2}{6 + 8w_1 + 2w_2}\punc,
	\end{aligned}
    \end{equation}
where \(w_1 = e^{-\Delta/T}\) and \(w_2 = e^{-4\Delta/T}\). These values are plotted as horizontal dashed lines in \reffig{mdensity}(a,b), and are in good agreement with the MC results.

As a final check that the system is reaching thermal equilibrium and that there is no dependence on the initial state, we compare with simulations starting from a random configuration (effectively \(T \gg \Delta\)). In this case, shown in \reffig{mdensity}(a,b) with dash--dot lines, the densities of both types of monopoles start at a large value and rapidly decrease, reaching an identical final density as with the field-quench protocol.

\subsection{Magnetization}
\label{secMagnetization}

In \reffig{mdensity}(c), we show the time evolution of the magnetization density, which decreases from its maximum value at \(t=0\) to zero at long times. We define
\begin{equation}
    m_z = \frac{M_z}{M\sub{sat}}\punc,
\end{equation}
which gives the \(z\) component of the magnetization relative to its saturation value \(M\sub{sat} = \lvert\hat{\boldsymbol{n}}_i \cdot \hat{\boldsymbol{z}}\rvert N\sub{s}\)
where \(\lvert\hat{\boldsymbol{n}}_i \cdot \hat{\boldsymbol{z}}\rvert = \frac{1}{\sqrt{3}}\) is the component of the fixed unit vector \(\hat{\boldsymbol{n}}_i\) along the \(z\) axis. The initial configuration is fully magnetized and so has \(m_z = 1\), while the full symmetry of the lattice is restored in the equilibrium configuration at long times, and so \(m_z = 0\).

As with the monopole densities, the relaxation is faster at higher temperatures. The dashed lines in \reffig{mdensity}(c) show exponential fits, \(m_z = e^{-t/\tau}\), for each temperature, and the fitted relaxation timescale \(\tau\) is plotted as a function of \(T\) in the left inset. The observed exponential growth of \(\tau\) with \(T^{-1}\) is consistent with previous studies using the standard model of spin ice dynamics \cite{Jaubert2011,Hallen2022}. The exponential fits match the MC data quite well for \(T \gtrsim \qty{3}{\kelvin}\), but become increasingly poor as \(T\) decreases. This is compatible with relaxation dominated by isolated spin flips at higher temperatures, with collective effects becoming significant only at lower \(T\). We discuss this early-time behavior in more detail in \refsec{SecEarlyTime}.

The right inset of \reffig{mdensity}(c) shows the same data plotted on a double-logarithmic vertical scale, so that a decrease of the form \(\sim e^{-(t/\tau)^\beta}\), would appear as a straight line with slope \(-\beta\). We see no indication of stretched-exponential decay, which would correspond to \(\beta < 1\), though the relaxation at the lowest temperatures is broadly compatible with a compressed exponential (\(\beta > 1\)).

\subsection{Finite-size effects and long-range interactions}

\begin{figure}
\includegraphics[width=0.85\linewidth]{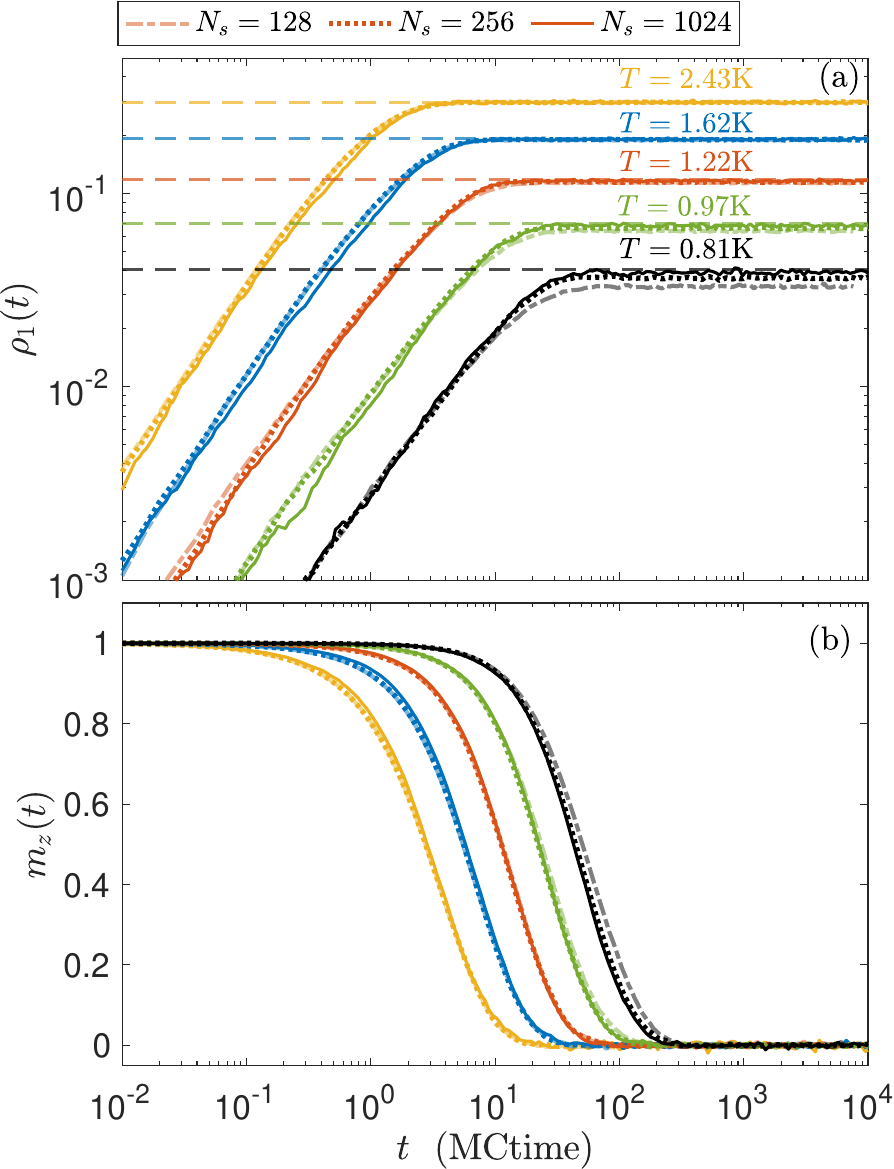}
\caption{Finite-size effects on relaxation of bulk properties: (a) Time evolution of density of (single) magnetic monopoles \(\rho_1\) for various system sizes \(N\sub{s}\) and temperatures \(T\). System dimensions are \(2\times2\times2\) (\(N\sub{s}=128\)), \(2\times2\times4\) (\(N\sub{s}=256\)) and \(4\times4\times4\) (\(N\sub{s}=1024\)) cubic unit cells. Dashed horizontal lines show the equilibrium monopole density in the thermodynamic limit for each \(T\). At the lowest \(T\) values, there is a significant finite-size effect in the equilibrium value of \(\rho_1\). (b) Time evolution of magnetization \(m_z(t)\) for the same parameters.}
\label{mdensity_FSE}
\end{figure}

\begin{figure}
\includegraphics[width=0.9\linewidth]{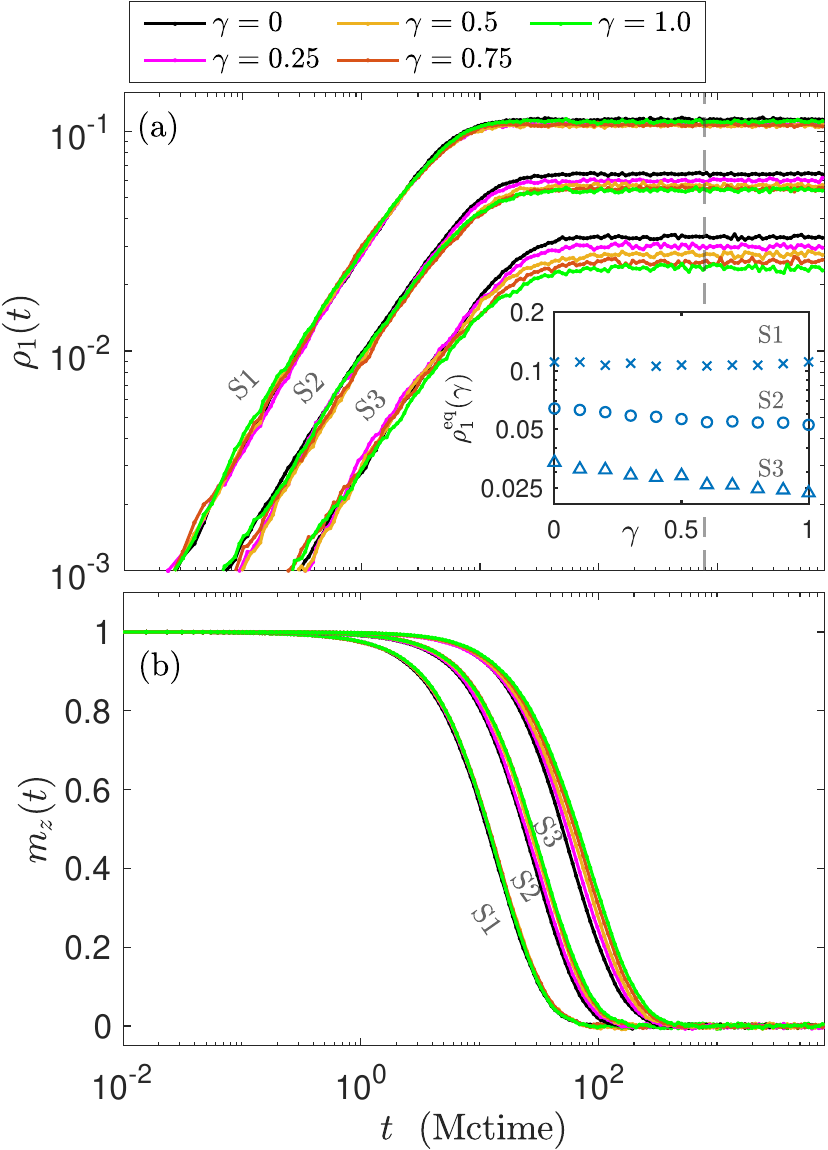}
\caption{Effects of long-range interactions on relaxation of bulk properties: (a) Time evolution of monopole density $\rho_1$ for various strengths of long-range interactions, where \(\gamma = 0\) has only nearest-neigbor interactions and \(\gamma=1\) is the dumbbell model (see \refsec{Sc:Hamiltonian} for details). The temperatures are \(T = \qty{1.22}{\kelvin}\) (S1), $T = \qty{0.97}{\kelvin}$ (S2), and $T = \qty{0.81}{\kelvin}$ (S3), and the system size is \(N\sub{s}=128\) in all cases. Inset: Long-time values $\rho_1\super{eq}$ of $\rho_1$, evaluated at the time corresponding to the vertical dashed line in the main figure, plotted versus $\gamma$. (b) Time evolution of magnetization $m_z$ for the same parameters.}
\label{cdensity_C}
\end{figure}

As a test of the sensitivity of our results to finite-size effects and long-range interactions, we show the dependence of monopole density and magnetization for various system sizes in \reffig{mdensity_FSE} and for nonzero \(\gamma\) in \reffig{cdensity_C}. In both cases, only small quantitative effects are seen.

The most noticeable effect of increasing \(\gamma\) from zero is a decrease in the equilibrium (i.e., long-time) density of monopoles \(\rho_1\), shown in the inset of \reffig{cdensity_C}(a). This occurs at all three temperatures shown, although for \(T=\qty{1.22}{\K}\) the slight decrease for small nonzero \(\gamma\) reverses as \(\gamma\) approaches \(1\). (Simulations at \(\gamma > 1\) [not shown] confirm that \(\rho_1\super{eq}\) increases at larger \(\gamma\) in all three cases.)

The decrease of monopole density with \(\gamma\) is, we believe, merely a consequence of the parameterization of the Coulomb interactions that we choose in our simulation: As \(\gamma\) increases, we also increase \(\nu\) in order to keep the activation energy \(\Delta\) fixed [see \refeq{Eq:Delta} and the following paragraph]. This has the effect of slightly increasing the energy of an oppositely charged pair of monopoles on tetrahedra beyond nearest neighbors, and hence reducing their density. At larger \(\gamma\), the longer-range interaction [second term in \refeq{eq:H}] can become larger, reversing the effect.

Increasing \(\gamma\) from zero also results in slower relaxation of the magnetization, as shown in Fig.~\ref{cdensity_C}(b). This effect is more pronounced at lower temperatures, and can be understood as a consequence of the reduced monopole density.

\section{String and cluster dynamics}

\label{Sc:SCdynamics}

Up to this point, we have been considering bulk properties, in terms of which the relaxation appears to be quite conventional. We now consider the microscopic processes by which this magnetization occurs, considering the basic processes that allow demagnetization of the fully saturated configuration.

\begin{figure}
\centering
\includegraphics[width=\columnwidth]{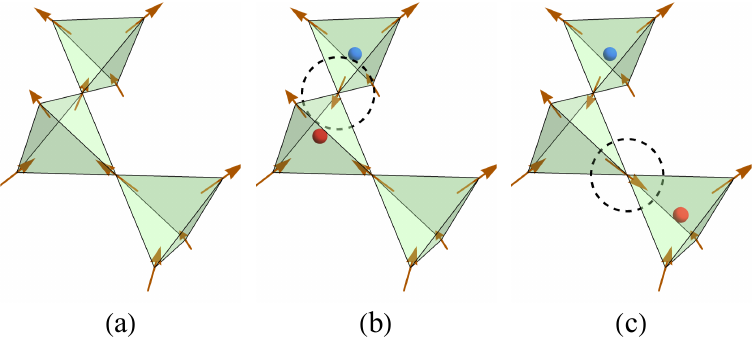}
\caption{Illustration of string formation and growth. (a) In the starting configuration, all spins point upwards, aligned with the external field applied at \(t<0\). (b) A spin flip produces a pair of monopoles on adjacent tetrahedra, which can be interpreted as a string of unit length, \(\ell = 1\). (c) Flipping a neighboring spin in the layer either above or (as shown) below the first moves one of the two monopoles, increasing the length of the string to \(\ell = 2\). From this configuration, the string could be further extended to length \(\ell = 3\), for example by flipping one of the two bottom-most spins. Note that flipping the top-right spin in the bottom tetrahedron would create a double monopole of charge \(n_\alpha = +2\), with a much higher energy cost. For this reason, an isolated string always follows the \(z\) direction and cannot form a closed loop (unless it spans the system boundaries).}
\label{FigString}
\end{figure}

The starting configuration has all spins polarized along the field and hence no monopoles. Starting from the polarized configuration, flipping any spin from up to down produces a pair of monopoles on neighboring tetrahedra, as illustrated in \reffig{FigString}, and so involves a large energy cost \(2\Delta\). Once such a pair has been produced, however, another process becomes possible: by flipping a second spin on either of the tetrahedra, its monopole can be moved to a neighboring tetrahedron; see \reffig{FigString}(c). The only energy cost associated with this second process is due to the change in the Coulomb interaction from separating the two monopoles, which is much smaller than \(\Delta\) (and zero in the case \(\gamma = 0\)).

This process can be continued, separating the two monopoles along the \(z\) direction, and leaving behind a string of downward-pointing spins \cite{Jaubert2008,Powell2008}, sometimes referred to as a Dirac string \cite{Morris2009}. Importantly, since these strings are defined with respect to the fully polarized initial configuration, an isolated string cannot form a closed loop unless it spans the periodic boundaries; an open string is always aligned along the \(z\) direction and has one monopole at each end.\footnote{Given a single spin configuration, there is no way to uniquely define the strings. In the context of the field quench, however, we can define the strings by reference to the initial configuration \cite{Morris2009}.}

The dynamics following the quench, at least at short times, can therefore be understood in terms of two processes: a slow process whereby a single spin is flipped, producing a pair of monopoles on neighboring tetrahedra; and a fast process where subsequent spin flips separate the monopoles and produce a string of flipped spins. On longer time scales, once a significant fraction of the spins have been flipped downward, the system is no longer well described by isolated strings but rather clusters of flipped spins.

\begin{figure}
\centering
\includegraphics[width=\columnwidth]{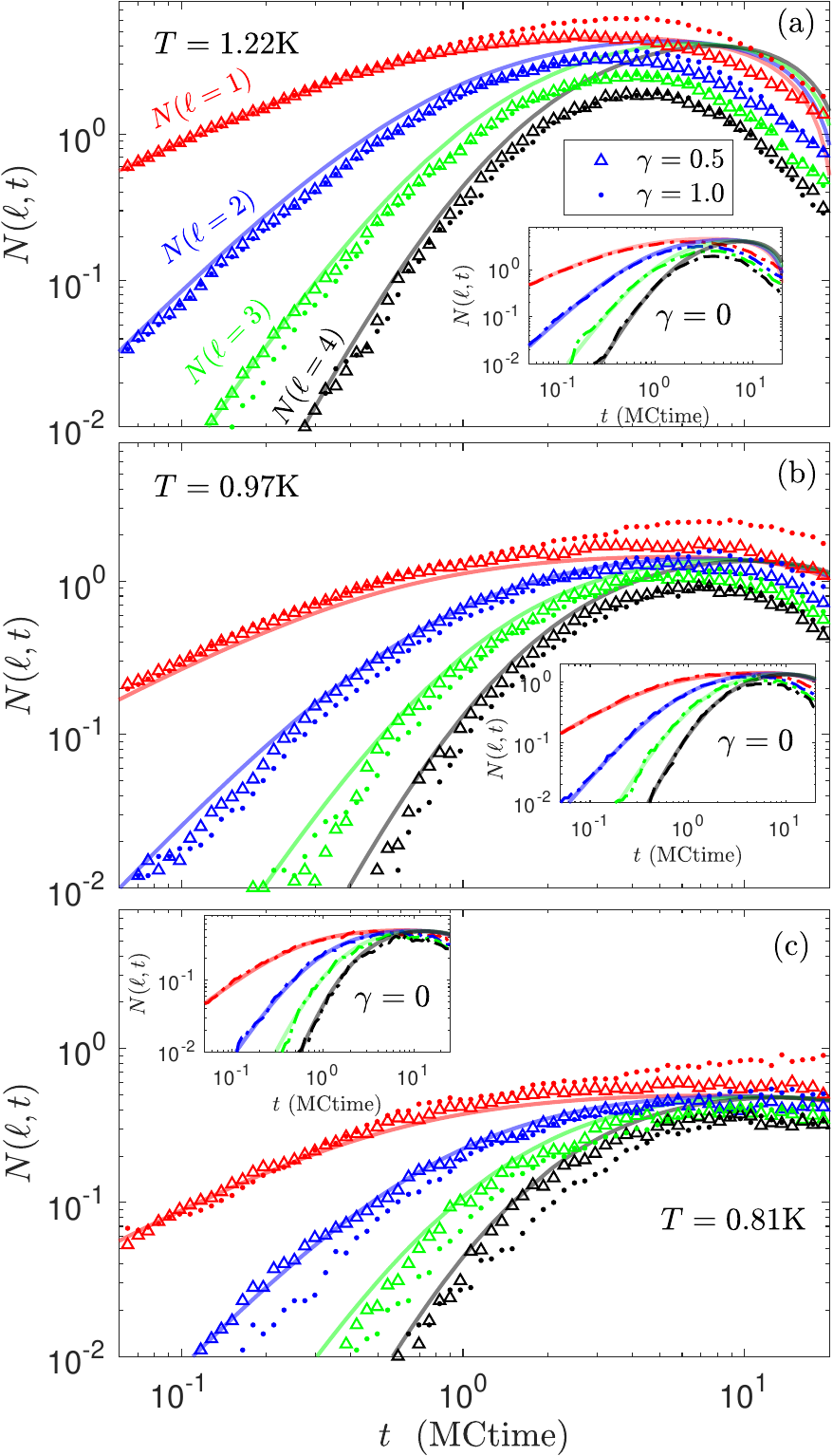}
\caption{Early-time dynamics: Mean number \(N(\ell,t)\) of connected sets of flipped spins of size \(\ell \le 4\) at time \(t\), for various temperatures.
In both main figures and insets, solid lines show the numerical solution of the single-string model defined in \refsec{SecSingleStringModel}, which assumes \(\gamma = 0\) and is valid only at early times. [We solve the coupled equations \refeqand{eq:upspins}{EqNellt} with a finite cutoff on \(\ell\), which is varied to confirm convergence.]
MC results, with \(N\sub{s}=1024\) spins (\(4\times4\times4\) cubic unit cells) and averaged over 500 independent runs, are shown with symbols in the main figures, where \(\gamma > 0\), and dashed lines in the inset, where \(\gamma = 0\).
Note that mean number is well below \(1\) at early times for most \(\ell\), indicating that most samples have no sets at all of that size.}
\label{stringshorttime}
\end{figure}
\subsection{Early-time dynamics: Formation of strings}
\label{SecEarlyTime}

To study the string--cluster crossover within our MC simulations, we identify, for each configuration, all connected sets of flipped (down) spins, where two spins are \emph{connected} if they are joined by a nearest-neighbor path of down spins. In \reffig{stringshorttime}, we show $N(\ell, t)$, the mean number of such sets of each size \(\ell \le 4\) at time \(t\) in a system with \(N\sub{s}=1024\) spins, for different temperatures \(T\) and dipolar interaction strengths \(\gamma\).

In all cases, \(N(\ell, t)\) grows at short times before decreasing, with successively larger \(\ell\) values growing more slowly at first. Comparing with \reffig{mdensity_FSE}, the maximum is seen to coincide roughly with the time at which the magnetization begins to decrease significantly from its initial value of \(m_z = 1\). We therefore interpret the initial increase as the regime where the strings are formed and grow independently. The decrease at later times indicates the crossover into cluster dynamics, which we address in \refsec{SecLateTimeDynamics}.

\subsubsection{Single-string model}
\label{SecSingleStringModel}

In the independent-string regime, it is possible to describe the dynamics analytically by considering the population of strings of each length \(\ell\) (in units of \(a\sub{d}\)). We assume that the density of down spins is low enough that we can treat each string as well isolated from all others, so that each string effectively grows in a background where all other spins point upwards. In addition, we assume that we can neglect finite-size effects, which is valid if all strings are much smaller than the system size in the \(z\) direction. We describe the \(\gamma = 0\) case first, and then comment briefly on the effect of long-range interactions.

\begin{figure}
\includegraphics{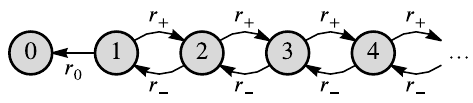}
\caption{Graph representing the dynamical model for a single string. Vertices (circles) denote strings of length \(\ell\), with \(\ell=0\) representing a string that has shrunk to zero length and hence disappeared. Edges (arrows) are transitions with associated rates for growth \(r_+=2\), contraction \(r_- = 1\), and annihilation \(r_0 = P\sub{G}(-2\Delta)\), in units of the spin-flip rate \(\tau\sub{flip}^{-1}\).}
\label{FigGraph}
\end{figure}

The number of strings of length \(\ell > 1\) changes due to growth and contraction of existing strings. In addition, we must consider creation and annihilation processes for strings of length \(\ell = 1\). To describe the dynamics, we define a stochastic model for a single string, illustrated in \reffig{FigGraph}: consider a string created at time \(t_0\) and let \(\Pr(\ell,s = t - t_0)\) be the probability that it has length \(\ell \in \mathbb{Z}_{\ge 0}\) at time \(t\). By definition, \(\Pr(\ell,0) = \delta_{\ell,1}\), while the only stationary state is \(\ell = 0\), an absorbing state representing a string that has shrunk to zero length and disappeared.

Growth of a string of length \(\ell\ge 1\), producing one of length \(\ell+1\), occurs when one flips any of four up spins, two in the row immediately above the top of the string and two in the row immediately below its bottom [see \reffig{FigString}(b)]. On the other hand, contraction of a string of length \(\ell > 1\) requires that one of two down spins, either the top-most or bottom-most of those comprising the string, should be flipped. For each of these processes, there is no change in energy (for \(\gamma = 0\)), because a monopole is effectively moved from one tetrahedron to another. Either update is therefore accepted with Glauber probability \(P\sub{G}(0) = \frac{1}{2}\). In units of the overall spin-flip rate \(\tau\sub{flip}^{-1}\), growth therefore occurs with rate \(r_+ = 4P\sub{G}(0) = 2\) and contraction with rate \(r_- = 2P\sub{G}(0) = 1\).\footnote{Note that an isolated string tends to grow, since \(r_+ > r_-\), and that this is due to entropy: there are more ways to grow a string than to shrink it. This effect is in agreement with entropic arguments about favorability of strings in equilibrium \cite{Jaubert2008}.}

Annihilation of a unit-length string is represented in the single-string model by the absorbing transition from \(\ell = 1\) to \(\ell = 0\). This occurs when the single down spin comprising the string is randomly chosen and flipped back upwards. This process reduces the energy by \(2\Delta\), where \(\Delta\) is the energy of a single monopole. The acceptance probability is therefore \(P\sub{G}(-2\Delta)\), and so the rate is simply \(r_0 = P\sub{G}(-2\Delta)\).

This stochastic model can be solved by expanding in terms of the eigenvectors of the rate matrix, as we describe in \refapp{AppSingleStringModel}. For the limiting case \(\Delta/T\rightarrow\infty\), where \(r_0 \rightarrow r_-\), the result can be expressed in terms of modified Bessel functions, while the general result can be expressed in closed form in terms of a contour integral, \refeq{EqPrltintegral}.

The single-string stochastic model could be modified to include the effect of long-range interactions, \(\gamma \neq 0\). In this case, changing the length of the string does involve an energy change, due to the Coulomb attraction between the monopoles at its ends. For \(\ell > 2\), the energy depends on the path of the string, rather than merely its length, and so one needs to distinguish all possible string shapes, with the number of states increasing exponentially with \(\ell\). The qualitative effect will be to reduce the rate at which short strings grow into longer ones, as seen in the numerical results in \reffig{stringshorttime}.

\subsubsection{String populations}
\label{SecStringPopulations}

A string of length \(\ell = 1\) is created when any spin is flipped from up to down, as long as all of its neighbors point upward. Since we assume that the density of down spins remains small, we approximate the number of sites where such a string can be created by the total number of up spins,
\begin{equation}
\label{eq:upspins}
N_\uparrow(t) = N\sub{s} - \sum_{\ell=1}^{\infty} \ell N(\ell, t)\punc,
\end{equation}
The process costs energy \(2\Delta\) and hence the spin flip is accepted with probability \(P\sub{G}(2\Delta)\). The rate at which strings are created at time \(t_0\) is therefore \(P\sub{G}(2\Delta)N_\uparrow(t_0)\), within this approximation. The string distribution at time \(t\),
\begin{equation}
\label{EqNellt}
N(\ell, t) = P\sub{G}(2\Delta)\int_0^t d t_0 \, N_\uparrow(t_0) \Pr(\ell, t-t_0)\punc,
\end{equation}
is then found by considering strings created at all previous times \(t_0\) and their probability of reaching size \(\ell\).

The behavior of \(N(\ell,t)\) to leading-order in \(t\) can be determined using \refeqand{EqNellt}{EqPrltleadingorder}. To this order, we set \(N_\uparrow = N\sub{s}\), replacing \refeq{eq:upspins}, and so the integral gives simply
\begin{equation}
\label{EqNelltearlytime}
N(\ell, t) \approx \frac{r_+^{\ell - 1}}{\ell!} N\sub{s}P\sub{G}(2\Delta) t^\ell
\end{equation}
to leading order in \(t\) for each \(\ell\).

Results obtained from the analytical model by numerical integration, with initial condition \(N(\ell,0) = 0\) for all \(\ell \ge 1\), are shown with solid lines in \reffig{stringshorttime}. (We apply a cutoff \(\ell\sub{max}\) on the maximum length \(\ell\) and check that the results are insensitive to the value of \(\ell\sub{max}\).) At very short times, they are in good quantitative agreement with those from the MC simulations for \(\gamma = 0\) (with no fitting parameters).

As expected, the results of the analytical model deviate significantly around the time where the maxima are reached and the assumptions cease to apply. As the MC results in \reffig{stringshorttime} show, the number of short strings decreases rapidly at larger times.

The model of independent strings necessarily fails when the magnetization falls well below its saturated value, and so the number of flipped spins can no longer be treated as small. In the terms of the number of up spins, \refeq{eq:upspins}, the magnetization density can be written as
\begin{equation}
    \label{Eqmz}
    m_z(t) = 2\frac{N_\uparrow(t)}{N\sub{s}} - 1= 1-\frac{2}{N\sub{s}}\sum_{\ell=1}^{\infty} \ell N(\ell, t)\punc.
\end{equation}

For \(T\) of order \(\Delta\), the rate of string creation is high and significant deviation from saturation magnetization occurs when a large number of strings has been created, even if most strings remain short. As \reffig{stringshorttime}(a) shows, even for fairly low temperature, \(T = \qty{1.22}{\kelvin} \simeq 0.44\Delta\) significant deviations from the single-string model occur while strings of length \(\ell = 1\) remain a clear majority.

On the other hand, for \(T \ll \Delta\), creation of strings is suppressed by \(P\sub{G}(2\Delta) \approx e^{-2\Delta/T}\), and string growth is the main process that reduces the magnetization. For intermediate times, where most strings are much longer than the lattice scale but shorter than their separation (and the system size), the single-string model is still valid. According to \refeq{EqMeanLengthLarget}, the mean string length is given in this regime by \(\sum_{\ell}\ell\Pr(\ell, t) \approx \frac{3}{2} + \frac{1}{2}t\), using \(P\sub{G}(-2\Delta)\simeq 1\) for \(T \ll \Delta\). Integrating over time as in \refeq{EqNellt} and again approximating \(N_\uparrow = N\sub{s}\) within the integral gives
\begin{equation}
m_z(t) \approx 1 - 2 P\sub{G}(2\Delta)\left(\frac{3}{2}t + \frac{1}{4}t^2\right)
\end{equation}
at early time. While this result gives a quantitatively good description of the data only at very short times, it provides a qualitative explanation of the deviation from exponential behavior, \(m_z = e^{-t/\tau}\), noted in \refsec{secMagnetization}.

\begin{figure*}
\includegraphics[width=0.85\linewidth]{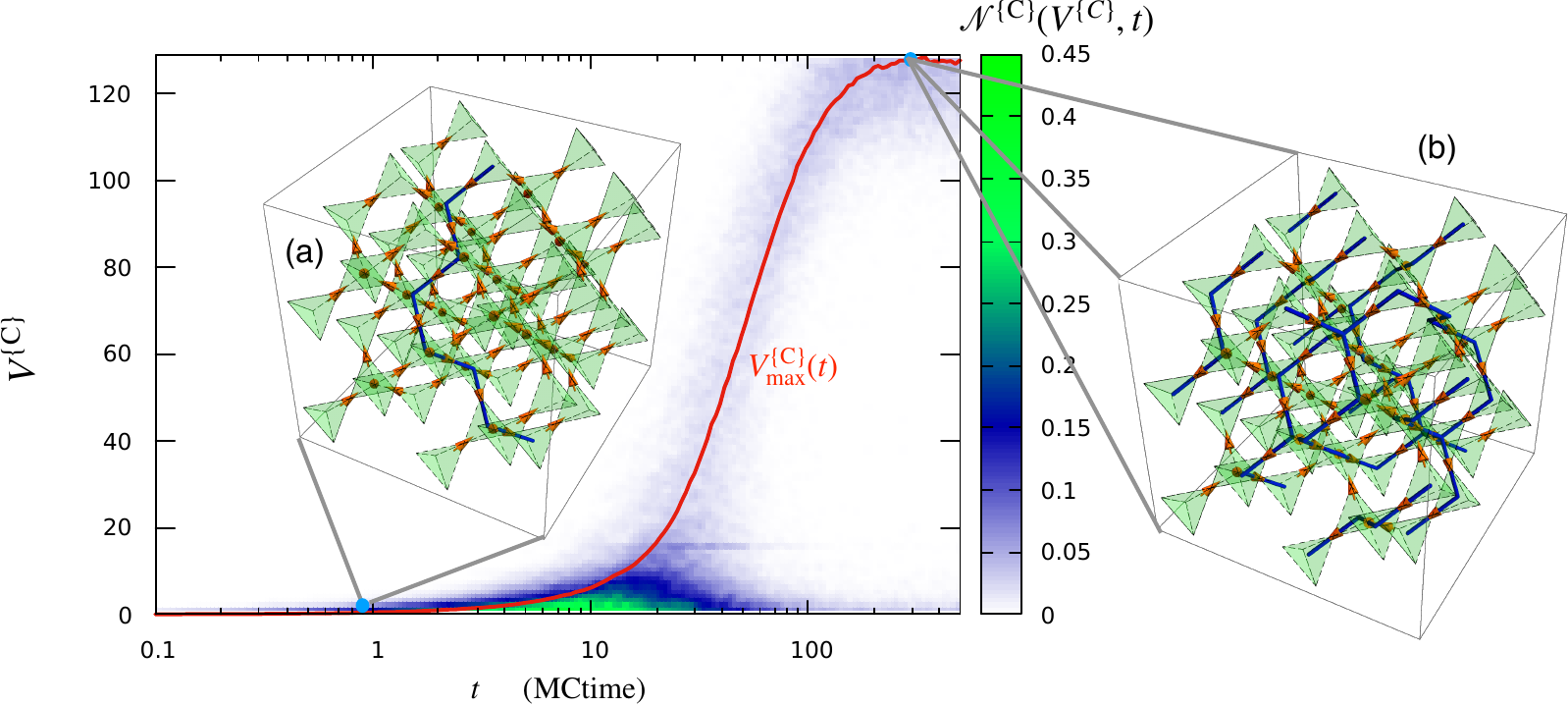}
\caption{Crossover from string to cluster dynamics following a field quench: Distribution of sizes \(V\sprC\) of connected sets of flipped spins across a sample of configurations, plotted on a color scale, as a function of time. The system contains \(N\sub{s}=256\) spins (\(2\times2\times4\) cubic unit cells) and has temperature \(T=\qty{0.81}{\kelvin}\). The red solid line shows the mean size \(V\sub{max}\sprC\) of the largest set, which reaches \(\sim N\sub{s}/2=128\) at late time. Insets illustrate typical configurations at early and late times, indicated with blue dots on the main figure. (A smaller system, with \(2\times2\times2\) cubic unit cells, is shown for clarity.) (a) Typical early-time configuration. A string of down (flipped) spins that spans the system is highlighted with a blue solid line; all other spins point upward. (b) Typical late-time configuration. Blue solid lines join the centers of neighboring tetrahedra containing flipped spins, which form a single large cluster filling nearly all of the system.}
\label{fig:SCtrans}
\end{figure*}

\subsection{Late-time dynamics: From strings to clusters}
\label{SecLateTimeDynamics}

While the short-time dynamics can be understood in terms of growth of isolated strings, this picture ceases to apply at longer times when the density of strings becomes large. In fact, on the pyrochlore lattice, strings cease to be uniquely defined at higher densities of flipped spins: where all four spins on a single tetrahedron point downwards, there are two equivalent choices of pairings that define different paths for the two strings. We therefore describe the dynamics at later times in terms of clusters of flipped spins; these may be viewed as dense networks of interwoven strings, with the caveat that there is no unique way to ``untangle'' the strings.

The crossover from strings to clusters is illustrated in \reffig{fig:SCtrans}(a) and (b), where connected sets of flipped spins are joined by solid blue lines. Inset (a) shows a typical configuration at early time, where a single string has grown along the \([100]\) direction (upwards), while inset (b) shows a late-time configuration containing a large cluster of flipped spins. In the latter case, the cluster includes nearly half of the spins in the lattice and is roughly isotropic, with no privileged orientation.

Using our MC algorithm, we produce a sample of independent runs at \(T=\qty{0.81}{\kelvin}\), and, at each time \(t\) in each run, identify all clusters (connected sets of flipped spins) in the configuration. The distribution of cluster volumes \(V\sprC\) (i.e., the number of clusters containing \(V\sprC\) flipped spins) is shown in \reffig{fig:SCtrans} (main figure). Superimposed on this (red solid line), we show the mean volume \(V\sub{max}\sprC\) of the largest cluster, with the average taken over the sample of runs.

\begin{figure}
\includegraphics[width=1.0\linewidth]{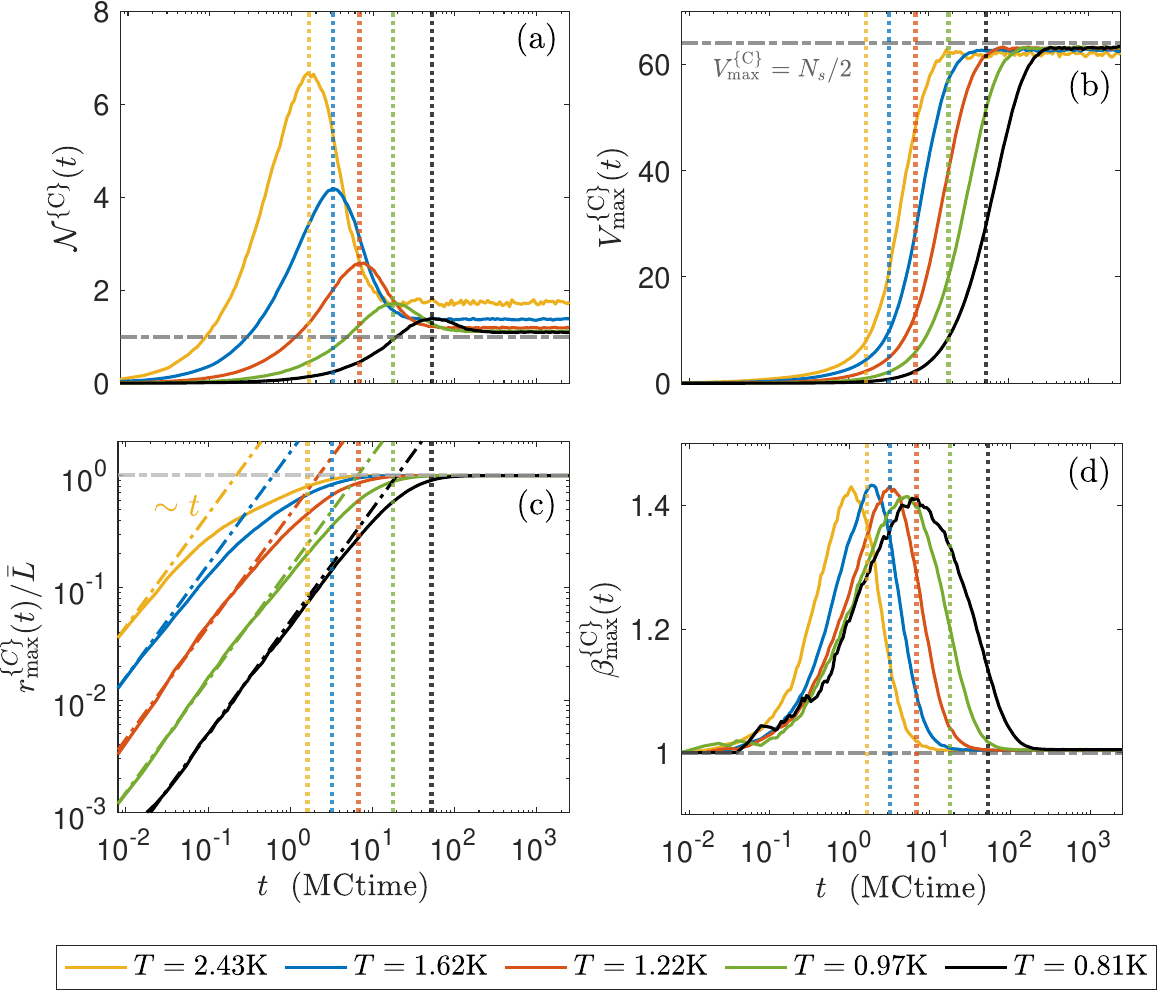}
\caption{Morphology of connected sets of flipped (i.e., downward-pointing) spins as a function of time \(t\), at different temperatures \(T\) for a system of \(N\sub{s}=128\) spins (\(2\times2\times2\) cubic unit cells). (a) Total number \(\mathcal{N}\sprC\) of connected sets of flipped spins. Dashed vertical lines (at the same times in each panel) mark the peak of \(\mathcal{N}\sprC\) at each \(T\). The dash--dot horizontal line is at \(\mathcal{N}\sprC = 1\). (b) Mean volume \(V\sub{max}\sprC\) (number of spins) of the largest such set. The dash--dot horizontal line is at \(V\sub{max}\sprC = N\sub{s}/2\), where half of the spins point downwards. (c) Linear size \(r\sub{max}\sprC\) of the largest set, divided 
by its maximum value, \(\bar{L}=\frac{1}{2\pi}\sqrt{L_x^2+L_y^2+L_z^2}\). (d) Aspect ratio \(\beta\sub{max}\sprC\) (see main text) of the largest set.}
\label{fig:string}
\end{figure}

We show in \reffig{fig:string}(a) and (b) respectively the total number \(\mathcal{N}\sprC\) of connected sets of flipped spins and the mean volume of the largest set \(V\sub{max}\sprC\) for several \(T\) values and \(N\sub{s}=128\) spins.

To measure the spatial extent of the largest set, we define
\begin{equation}
\label{eq:definermaxC}
r\sub{max}\sprC = \sqrt{\big(x\sub{max}\sprC\big)^2 + \big(y\sub{max}\sprC\big)^2 + \big(z\sub{max}\sprC\big)^2}\punc,
\end{equation}
where
\begin{equation}
\label{eq:definexmaxC}
\big(x\sub{max}\sprC\big)^2
= \left(\frac{L_x}{2\pi}\right)^2\left(1 - \left\lvert \frac{1}{N\sub{d}\sprC}\sum_\alpha e^{2\pi i x_\alpha/L_\mu} \right\rvert^2\right)\punc,
\end{equation}
and similarly for \(y\sub{max}\sprC\) and \(z\sub{max}\sprC\). In \refeq{eq:definexmaxC}, the sum runs over the \(N\sub{d}\sprC\) tetrahedra \(\alpha\), with centers at \(\boldsymbol{r}_\alpha = (x_\alpha, y_\alpha, z_\alpha)\), that contain at least one flipped spin belonging to the largest connected set. This definition of \(r\sub{max}\sprC\) reduces to the root-mean-square radius (rms distance from the centroid) for a small set not spanning the system boundaries, but correctly accounts for the periodic boundary conditions, depending on \(x_\alpha\) only modulo \(L_x\). It saturates at \(r\sub{max}\sprC = \bar{L} \equiv \frac{1}{2\pi}\sqrt{L_x^2+L_y^2+L_z^2}\) for a cluster that fills the system uniformly.

In \reffig{fig:string}(c) and (d) respectively, we plot, as functions of time, \(r\sub{max}\sprC\) and the aspect ratio \(\beta\sub{max}\sprC = z\sub{max}\sprC/x\sub{max}\sprC\), a measure of the anisotropy of the largest set. When this forms a long isolated string, such as shown in \reffig{fig:SCtrans}(a), \(\beta > 1\), while for a cluster that fills the (isotropic) system, such as \reffig{fig:SCtrans}(b), \(\beta\simeq 1\). For the smallest possible set, where a single flipped spin shared by two adjacent tetrahedra, \(x\sub{rms} = z\sub{rms} \simeq a_d/2\), and so \(\beta = 1\).

Considering the behavior of these quantities, the dynamics can be divided into three stages. The initial rise in \(\mathcal{N}\sprC\) is due to the proliferation of isolated short strings, as discussed in \refsec{SecEarlyTime}. For all temperatures, \(r\sub{max}\sprC\) grows linearly with \(t\) at the earliest times and the anisotropy measure \(\beta\sub{max}\sprC\) increases rapidly from \(1\), consistent with linear growth of isolated strings. For small \(T\), this growth of \(r\sub{max}\sprC\) continues until it reaches a value of order the linear system size.

The process of string formation is suppressed at low temperatures, since it involves the creation of monopole pairs. The peak in \(\mathcal{N}\sprC\) is therefore lower and occurs at later time for lower \(T\). For all temperatures, this peak is reached at roughly the time when largest set begins to occupy a significant fraction of the system, with \(V\sub{max}\sprC\) of order \(10\%\) of \(N\sub{s}\).

The subsequent decrease of \(\mathcal{N}\sprC\) can be interpreted as a consolidation process, whereby strings merge into longer strings and clusters. A pair of strings can join if the monopoles of opposite charge on their ends reach the same tetrahedron and hence annihilate. If instead a monopole enters a tetrahedron through which another string passes, the result is a single cluster that can no longer be viewed in terms of isolated strings.

In the third stage, at long times, \(\mathcal{N}\sprC\) reaches a constant value that approaches \(1\) at low temperature, while \(V\sub{max}\sprC\) saturates at \(\frac{1}{2}N\sub{s}\). In this regime, a single large cluster fills the system, as illustrated in \reffig{fig:SCtrans}(b), containing approximately half of the spins. Such a cluster has linear extent \(r\sub{max}\sprC\) equal to its maximum value, \(\bar{L}\), and is approximately isotropic, giving \(\beta\sub{max}\sprC = 1\), as seen at late times in \reffig{fig:string}(c) and (d).

\begin{figure}
\includegraphics[width=0.8\linewidth]{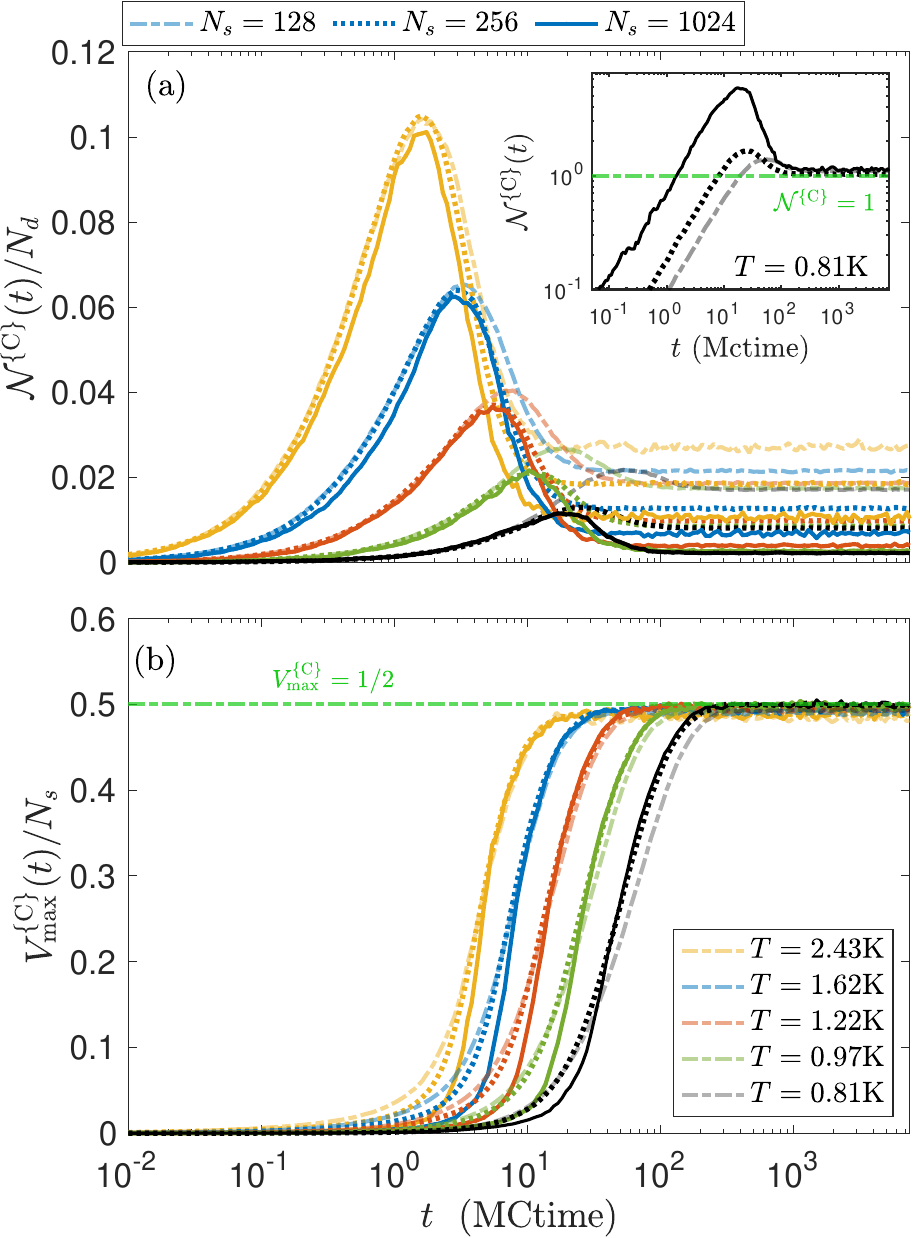}
\caption{Finite-size effects on the string--cluster crossover: (a) Number \(\mathcal{N}\sprC\) of connected sets of flipped spins versus time for various system sizes (line styles; same dimensions as in \reffig{mdensity_FSE}) and temperatures (colors). In the main figure, \(\mathcal{N}\sprC\) is scaled by the system size (specifically, the number of tetrahedra \(N\sub{d}\)). The inset shows the data for \(T = \qty{0.81}{\kelvin}\) without scaling, which indicates that \(\mathcal{N}\sprC\rightarrow1\) at long time. (b) Mean volume \(V\sub{max}\sprC\) of largest connected set of flipped spins, scaled by the number of pyrochlore lattice sites \(N\sub{s}\).}
\label{stringNC_FSE2}
\end{figure}

To investigate the consequences of finite-size effects, we show results for different system sizes in \reffig{stringNC_FSE2}. At early time, the number \(\mathcal{N}\sprC\) of connected sets of flipped spins is approximately proportional to system size (number of tetrahedra) \(N\sub{d}\), consistent with independent formation and growth of strings. For the higher temperature values, \(\mathcal{N}\sprC\) continues to scale with \(N\sub{d}\) up to and somewhat beyond its peak, indicating that consolidation of clusters dominates over formation once they reach a certain (\(T\)-dependent, but \(N\sub{d}\)-independent) density.

By contrast, for the lowest temperature [\(T=\qty{0.81}{\kelvin}\); see inset of \reffig{stringNC_FSE2}(a)], the maximum value of \(\mathcal{N}\sprC\) grows more slowly than \(N\sub{d}\). This suggests that in this case finite-size effects on individual strings are already important at the start of the consolidation process.

At late time, the number of sets \(\mathcal{N}\sprC\) instead approaches \(1\) for all system sizes, with a single large cluster of down spins filling the system.

 \begin{figure*}
\includegraphics[width=\linewidth]{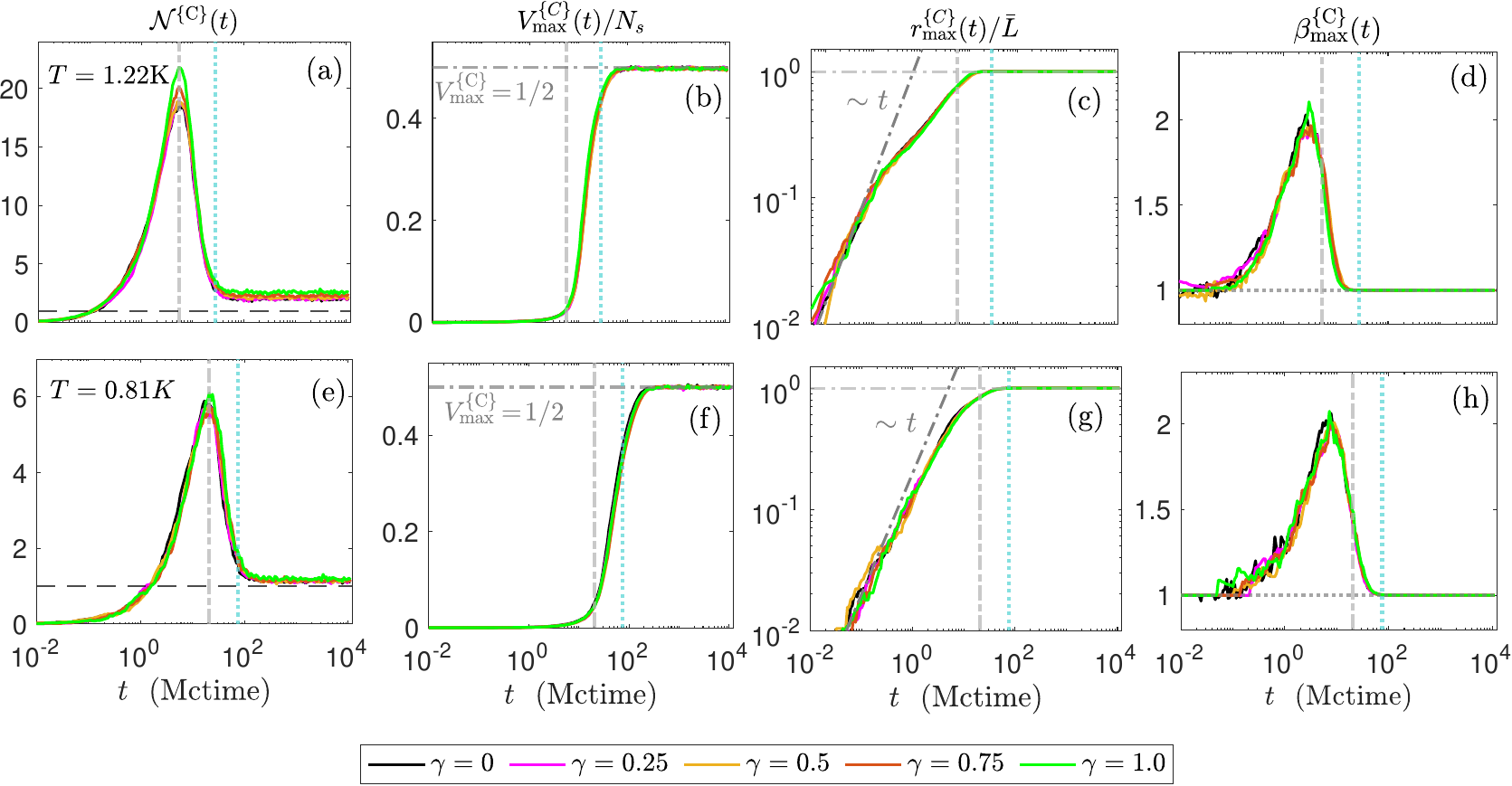}
\caption{Effects of long-range interactions on the string--cluster crossover: Same quantities as in \reffig{fig:string} for various strengths of long-range interactions \(\gamma\) (see \reffig{cdensity_C}). The top and bottom rows have \(T = \qty{1.22}{\kelvin}\) and \(\qty{0.82}{\kelvin}\) respectively, and \(N\sub{s}=1024\) spins (\(4\times4\times4\) cubic unit cells) in both cases. Dashed vertical lines (at the same times in each panel) mark the peak of \(\mathcal{N}\sprC\) at each \(T\).}
\label{stringC1024}
\end{figure*}

Results including long-range interactions are shown in \reffig{stringC1024}. Comparison with the case \(\gamma = 0\) indicates that there is no significant change in the qualitative behavior, and indeed little quantitative change. Coulomb interactions between monomers are expected to suppress the initial growth of strings, as noted in \refsec{SecEarlyTime}, and may also favor reconnection of long strings at the expense of cluster formation.

\section{Percolation}
\label{Sc:perco}

\begin{figure*}
\includegraphics{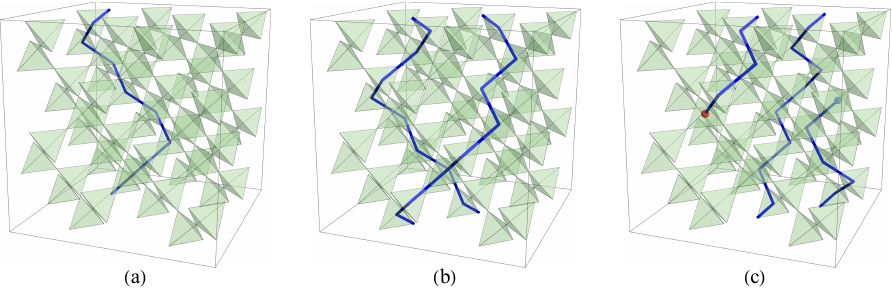}
\caption{Examples of criterion for percolation. Blue lines represent strings of flipped spins relative to the fully polarized configurations. (a,b) Configurations in which a percolating string exists. (c) A configuration where no such string exists. A set of flipped spins is considered to percolate the system only if it is closed and spans the periodic boundaries in the \([100]\) direction, shown upwards. This excludes cases such as (c), where an open string spans the system, but cannot be assigned a nonzero winding number.}
\label{FigPercolation}
\end{figure*}

Finally, we consider the growth of clusters from the perspective of percolation theory \cite{Stauffer2010, Lorenz1998}. Roughly speaking, we consider the set of flipped spins to be percolating if there exists a cluster that spans the periodic boundaries in the \([100]\) direction. Examples of this criterion are shown in \reffig{FigPercolation}; note that we choose to include only cases where a cluster has nontrivial winding numbers and exclude open strings. (The motivation for this choice is that it corresponds to the criterion for the equilibrium Kasteleyn transition in the limit \(\Delta/T \rightarrow \infty\) \cite{Jaubert2008}, and therefore provides a way to extend this to quench dynamics.)

\begin{figure}
\includegraphics[width=0.9\linewidth]{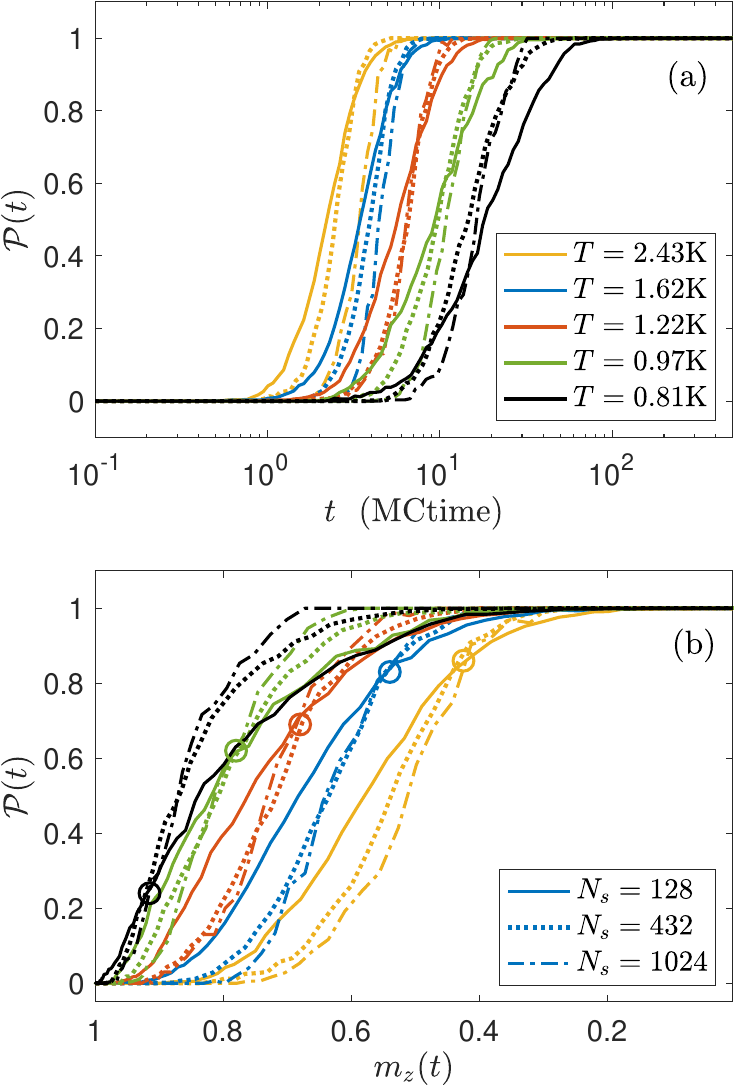}
\caption{Percolation probability \(\mathcal{P}\) versus (a) time \(t\) and (b) magnetization density \(m_z(t)\). System dimensions (shown with line styles) are \(2\times2\times2\) (\(N\sub{s}=128\)), \(3\times3\times3\) (\(N\sub{s}=432\)) and \(4\times4\times4\) (\(N\sub{s}=1024\)) cubic unit cells. For each temperature \(T\) (colors) a circle is used in (b) to highlight an apparent crossing point for different \(N\sub{s}\). These crossing points suggest the possibility of a continuous percolation transition.}
\label{percoNC}
\end{figure}

In \reffig{percoNC}(a), we show the probability \(\mathcal{P}\) that a percolating cluster exists (i.e., the fraction of samples that contain such a cluster), as a function of time \(t\) for various system sizes \(N\sub{s}\). The same data are shown in \reffig{percoNC}(b) but plotted as a function of magnetization \(m_z\). (These results are all for the case without long-range interactions, \(\gamma = 0\).) In both cases we see a crossover from \(\mathcal{P} = 0\) to \(1\), which becomes sharper as the system size increases.

In the limit \(T/\Delta\rightarrow\infty\), where the spins are independent and flipped randomly, this process is identical to standard bond percolation \cite{Scullard2020, Martins2003} on the diamond lattice, with occupation probability \(p = N_\downarrow/N\sub{s} = \frac{1}{2}(1-m_z)\). (Our model has spins on the sites of pyrochlore lattice, which are equivalent to the links of the diamond lattice.) We therefore expect a percolation transition at \(p_c = 0.39\) \cite{Gaunt1978,Galam1996,Stauffer2010}, manifesting as a crossing in \(\mathcal{P}\) as a function of \(m_z\) at \((m_z)\sub{c} = 0.22\), becoming sharper with larger system size.

For lower temperatures, we expect strings to percolate at lower \(p\) for any given system size, and hence higher \(m_z\), than independent flipped spins. (As an example, the configuration shown in \reffig{fig:SCtrans}(a) has a single string, and hence a low density of flipped spins, but is nonetheless percolating.) In our simulation results, we indeed find that decreasing \(\Delta/T\) causes each curve of \(\mathcal{P}\) to shift towards higher \(m_z\). In fact, we see evidence that the crossing point characteristic of a continuous transition remains for all \(\Delta/T\).

\section{Conclusions and Outlook}
\label{Sc:conclusions}

In this work, we have used MC simulations to study the dynamics of classical spin ice following a magnetic-field quench, where a strong field along the [100] crystal direction is suddenly removed. We have shown how the early-time dynamics can be understood through the formation and growth of strings of flipped spins, terminated at each end by magnetic monopoles, and presented exact results for a simple analytical model of this process. During this stage, the magnetization relaxes rapidly, though with significant deviations from a simple exponential decay for \(T \lesssim \qty{2}{\kelvin}\), and the monopole density increases from zero towards its equilibrium value.

At longer times, the system approaches a completely demagnetized equilibrium state. This can be interpreted as a process where the strings merge into clusters that eventually form a network extending throughout the whole system. We have also provided evidence for a percolation transition as a function of magnetization, which extends the equilibrium Kasteleyn transition \cite{Jaubert2008} to the dynamics. Throughout our results, we find that finite system size and dipolar spin--spin interactions (incorporated here using the dumbbell approximation) have relatively modest effects.

This work is partly motivated by experimental realizations of field quenches in spin ice materials \cite{Paulsen2014} that have been performed using \DTO\ in fields along both the [100] and [111] crystal directions. Our results are in general qualitative agreement with these experiments, which observed  relaxation of the magnetization with a timescale increasing rapidly with decreasing temperature \cite{Paulsen2014}. (They also observed only minor effects from dipolar interactions, finding good agreement with nearest-neighbor simulations.)

One important confounding factor for quantitative comparisons is the question of the extent to which the system remains in thermal equilibrium. Our simulations make the significant simplification of using Glauber dynamics with a fixed temperature, in effect assuming that thermal energy is always transferred rapidly enough for the system to remain locally in thermal equilibrium. In reality, this is certainly not always the case, as evidenced by occurrence of magnetic avalanches induced by local heating \cite{Fennell2005,Jackson2014}. Incorporating these effects into the theoretical framework developed here is a challenging problem that we defer to future work.

Our simulations use the so-called ``standard model'' of spin ice dynamics \cite{Ryzhkin2005,Jaubert2009}, where spin flips are attempted at a single fixed rate and accepted with a temperature-dependent probability. Recent work \cite{Tomasello2019,Hallen2022} has found that the low-temperature relaxation is better described by including two distinct flip rates, reflecting a bimodal distribution of local field strengths.

The consequences of this for the quench dynamics described here can be inferred at early times: For an isolated string, growth and contraction always involves flipping a spin in an asymmetric environment (see \reffig{FigString}), which occurs at the faster rate. By contrast, creation and annihilation of unit-length strings occurs in a symmetric environment and hence at the slower rate. The main result at early times is therefore an effective renormalization of the ``seeding'' rate to a smaller value. At later times the string density is higher, making the effects more difficult to predict, and simulations incorporating the two rates are required.

Future work will study quenches where the final magnetic field strength is nonzero, including close to the Kasteleyn transition and to the low-temperature side.

\acknowledgements
This work was supported by the Engineering and Physical Sciences Research Council grant number EP/T021691/1. The simulations used resources provided by the University of Nottingham High-Performance Computing Service.
\appendix
	  
\section{Algorithm to identify a percolated network of flipped spins}

At any time $t$, we consider each of the independent flipped spin networks in turn to find a percolated network. We consider all the diamond lattice sites which are the part of a particular flipped spin network and call this set $\{\mathcal{D}_i\}$. We save all these points in $\{\mathcal{D}_i\}$ for future use. The goal is to find out whether the network is percolated using the following algorithm. 
 

\begin{itemize}
    \item Step 1: Consider all the diamond lattice sites which are at the bottom boundary ($z=0$) of the cubic box. These sites (points) are the starting point of our numerical inspection to find probable percolating networks. Keep those starting points ($\{\mathcal{S}_0^i\}$) and their coordinates stored. 
    
    \item Step 2: Choose a random starting point $S_0 \in \{\mathcal{S}_0^i\}$. 

    \item Step 3: Check the necessary condition ($\mathcal{C}_1$) --- Does any of the two nearest neighbours (whose $z$ coordinates are higher than that of $S_0$) of $S_0$, belong to $\{\mathcal{D}_i\}$? The answer will be one of the following three scenarios: i) Only a single nearest neighbour (say $S_1$) satisfies $\mathcal{C}_1$, referring the direction [$S_0\rightarrow S_1$] to be the probable path of percolation. ii) Both the nearest neighbours (say $S_1$ and $S'_1$) satisfy $\mathcal{C}_1$ which indicates two possible paths of percolation (branching). In this case, we consider the same direction [$S_0\rightarrow S_1$] as in (i). But, along with it, we also keep $S'_1$ stored in a stack.
    iii) $\mathcal{C}_1$ is not satisfied. We change the starting point $S_0$ and repeat step 3 to find $S_1$. Keep searching for all starting points $S_0$. If $S_1$ is never found, we conclude the system is unpercolated.

    \item Step 4: Repeat step 3 for the point $S_1$ instead of $S_0$. It can update the direction [$S_0\rightarrow S_1 \rightarrow S_2$] if $\mathcal{C}_1$ is satisfied. Also for branching, like step 3(ii), we consider the path [$S_0\rightarrow S_1 \rightarrow S_2$] while updating the stack with [$S'_1\rightarrow S'_2$]. 

    \item Step 5: Repeat step 4. Again the following two scenarios may occur: (i) For any arbitrary $n$, if eventually we find $S_n = S_0$ (requirement $\mathcal{R}^{\ast}$), we conclude that the network [$S_0\rightarrow S_1 \rightarrow S_2 \rightarrow \cdots \rightarrow S_n$] percolates. (ii) If for any arbitrary $p$ step it fails to find $S_p$, then change the direction of search by choosing $S'_{r}$, where $S'_{r}$ is the latest point of the stack [$S'_1 \rightarrow S'_2 \rightarrow \cdots \rightarrow S_r'$]. Keep updating the path along this branch according to step 4. The idea is that, if at any stage the algorithm fails to update the direction along a specific branch, it chooses alternative branch by coming back to the latest primed point in the stack.

    \item Step 6: If steps 2--5 fails to find any percolated path, we change the starting point. That is, we choose another $S_0 \in \{\mathcal{S}_0^i\}$ and repeat from step 2. By choosing all the $S_0\in \{\mathcal{S}_0^i\}$ in turn, if $\mathcal{R}^{\ast}$ is never satisfied, the algorithm concludes that no percolation exists for the considered flipped spin network.
    
    \item Step 7: Repeat steps 1--6 for other flipped spin networks. For any chosen network, if the percolated path is found then algorithm concludes that the system has percolated at $t$. Otherwise, the system is unpercolated at time $t$. 
    
    \end{itemize}

\begin{widetext}

\section{Solution of single-string model}
\label{AppSingleStringModel}

The rate matrix for the stochastic process defined in \reffig{FigGraph} can be written in Dirac notation as
\begin{equation}
W = -(r_0 - r_-)\lvert 1\rangle\langle 1 \rvert + \sum_{\ell=1}^{\infty}\left[-(r_-+r_+)\lvert \ell\rangle\langle \ell\rvert + r_+\lvert \ell+1 \rangle \langle \ell \rvert + r_- \lvert \ell \rangle \langle \ell+1 \rvert\right]\punc,
\end{equation}
where we omit the absorbing state \(\lvert 0 \rangle\) (so the probability vectors will not be normalized).

This matrix has eigenvectors
\begin{equation}
\lvert \psi(z) \rangle = \sum_{\ell=1}^{\infty} \lvert \ell \rangle \left[ z^\ell - \frac{r_+ + (r_0 - r_-)z}{r_0 - r_- + r_-z}\left(\frac{r_+}{r_-z}\right)^{\ell-1} \right]
\end{equation}
with eigenvalues
\begin{equation}
\lambda(z) = r_+(z^{-1}-1) + r_-(z - 1)\punc.
\end{equation}

One can therefore write
\begin{equation}
\lvert 1 \rangle = -\oint_{\mathfrak{C}} \frac{d z}{2\pi i} \frac{1}{z}\frac{r_0 - r_- + r_-z}{r_+ + (r_0 - r_-)z}\lvert\psi(z)\rangle\punc,
\end{equation}
where the contour \(\mathfrak{C}\) encloses the origin (in a counterclockwise direction) but not the pole at \(z = r_+/(r_- - r_0)\). This implies, for any \(\ell \ge 1\),
\begin{align}
\Pr(\ell,t) &= \langle \ell \rvert e^{Wt} \lvert 1 \rangle \\
&= -\oint_{\mathfrak{C}} \frac{d z}{2\pi i} \frac{1}{z}\frac{r_0 - r_- + r_-z}{r_+ + (r_0 - r_-)z}e^{t\lambda(z)}\left[ z^\ell - \frac{r_+ + (r_0 - r_-)z}{r_0 - r_- + r_-z}\left(\frac{r_+}{r_-z}\right)^{\ell-1} \right]\\
&= \oint_{\mathfrak{C}} \frac{d z}{2\pi i} z^{\ell-2}\frac{r_+ - r_- z^2}{r_+ + (r_0 - r_-)z}e^{t[r_+(z^{-1}-1) + r_-(z - 1)]}
\punc,
\label{EqPrltintegral}
\end{align}
with the same conditions on the contour. (The substitution \(z' = z^{-1}r_+/r_-\) has been used in the second term in the integrand.) As described in \refsec{SecStringPopulations}, this integral can be performed numerically to find \(\Pr(\ell, t)\) and hence the string distribution \(N(\ell,t)\).

\subsection{Mean string length at late time}

The mean length of the string at time \(t\) is
\begin{equation}
\bar{\ell}(t) = \sum_{\ell = 1}^{\infty} \ell \Pr(\ell,t) = I(\mathfrak{C})\punc,
\end{equation}
where, using \refeq{EqPrltintegral},
\begin{equation}
I(\mathfrak{C}) = \oint_{\mathfrak{C}} \frac{d z}{2\pi i} \frac{1}{z} \frac{1}{(z-1)^2}\frac{r_+ - r_- z^2}{r_+ + (r_0 - r_-)z}e^{t[r_+(z^{-1}-1) + r_-(z - 1)]}\punc,
\end{equation}
and the contour \(\mathfrak{C}\) must be restricted to  \(\lvert z \rvert < 1\).
To determine the large-\(t\) behavior, consider a circular contour \(\mathfrak{C}'\) of radius \(\lvert z\rvert = \sqrt{{r_+}/{r_-}} > 1\). Since \(\mathfrak{C}'\) encloses a pole at \(z=1\), the integrals around the two contours are related by
\begin{align}
I(\mathfrak{C}') &= I(\mathfrak{C}) + \operatorname{Res}\left(\frac{1}{z} \frac{1}{(z-1)^2}\frac{r_+ - r_- z^2}{r_+ + (r_0 - r_-)z}e^{t[r_+(z^{-1}-1) + r_-(z - 1)]}, z=1\right)\\
&= I(\mathfrak{C}) - \frac{r_+(r_+-r_-+2r_0)}{(r_+-r_-+r_0)^2} - \frac{(r_+-r_-)^2}{r_+-r_-+r_0}t
\punc,
\end{align}
where \(\operatorname{Res}\) denotes the residue. The contour \(\mathfrak{C}'\) passes through stationary points of \(\lambda(z)\) at \(z = \pm\sqrt{{r_+}/{r_-}}\) and so can be evaluated for large \(t\) using the saddle-point approximation; the result decreases exponentially with \(t\) because \(\lambda\left(\pm\sqrt{{r_+}/{r_-}}\right) < 0\).

The dominant behavior of \(I(\mathfrak{C})\) at long time is therefore simply given by the residue of the pole at \(z=1\),
\begin{equation}
\label{EqMeanLengthLarget}
\bar{\ell}(t) \approx \frac{(r_+-r_-)^2}{r_+-r_-+r_0}t + \frac{r_+(r_+-r_-+2r_0)}{(r_+-r_-+r_0)^2}
\end{equation}
for large \(t\).

\subsection{Early time}

Using the integral representation, \refeq{EqPrltintegral}, one can also find the contribution to \(\Pr(\ell,t)\) of leading order in \(t\) for each \(\ell\). This is given by the lowest power of \(t\) that contributes to the residue of the pole at \(z=0\), which comes from the \((\ell-1)\)th term in the Taylor expansion of the exponential. The result is
\begin{equation}
\label{EqPrltleadingorder}
\Pr(\ell,t)\approx\frac{t^{\ell-1}r_+^{\ell-1}}{(\ell-1)!}\punc,
\end{equation}
to leading order in \(t\) for each \(\ell\ge 1\).

\subsection{Zero-temperature limit}

As described in the main text, the contraction and annihilation rates are given by \(r_-=1\) and \(r_0 = P\sub{G}(-2\Delta)\) respectively. For small \(T/\Delta\), we have \(P\sub{G}(-2\Delta)\simeq 1\) and so these rates are approximately equal. In the limit \(T/\Delta\rightarrow0\), where they are exactly equal, \refeq{EqPrltintegral} can be expressed in terms of the modified Bessel function \(\mathcal{I}_m\) \cite{NIST:DLMF:10}, defined by
\begin{equation}
\mathcal{I}_m(q) = \oint \frac{d z}{2\pi i} z^{-m-1}e^{\frac{1}{2}q(z+z^{-1})}\punc.
\end{equation}
Comparing with \refeq{EqPrltintegral}, this gives
\begin{align}
\Pr(\ell, t) &= e^{-t(r_++r_-)}\left(\frac{r_+}{r_-}\right)^{\frac{\ell-1}{2}} \left[\mathcal{I}_{\ell-1}\left(2t\sqrt{r_+r_-}\right)-\mathcal{I}_{-\ell-1}\left(2t\sqrt{r_+r_-}\right)\right]\\
&=e^{-t(r_++r_-)}\left(\frac{r_+}{r_-}\right)^{\frac{\ell-1}{2}} \frac{\ell}{t\sqrt{r_+r_-}}\mathcal{I}_{\ell}\left(2t\sqrt{r_+r_-}\right)
\punc.
\end{align}

\end{widetext}

\bibliography{spinice}{}

\begin{thebibliography}{58}%
\makeatletter
\providecommand \@ifxundefined [1]{%
 \@ifx{#1\undefined}
}%
\providecommand \@ifnum [1]{%
 \ifnum #1\expandafter \@firstoftwo
 \else \expandafter \@secondoftwo
 \fi
}%
\providecommand \@ifx [1]{%
 \ifx #1\expandafter \@firstoftwo
 \else \expandafter \@secondoftwo
 \fi
}%
\providecommand \natexlab [1]{#1}%
\providecommand \enquote  [1]{``#1''}%
\providecommand \bibnamefont  [1]{#1}%
\providecommand \bibfnamefont [1]{#1}%
\providecommand \citenamefont [1]{#1}%
\providecommand \href@noop [0]{\@secondoftwo}%
\providecommand \href [0]{\begingroup \@sanitize@url \@href}%
\providecommand \@href[1]{\@@startlink{#1}\@@href}%
\providecommand \@@href[1]{\endgroup#1\@@endlink}%
\providecommand \@sanitize@url [0]{\catcode `\\12\catcode `\$12\catcode
  `\&12\catcode `\#12\catcode `\^12\catcode `\_12\catcode `\%12\relax}%
\providecommand \@@startlink[1]{}%
\providecommand \@@endlink[0]{}%
\providecommand \url  [0]{\begingroup\@sanitize@url \@url }%
\providecommand \@url [1]{\endgroup\@href {#1}{\urlprefix }}%
\providecommand \urlprefix  [0]{URL }%
\providecommand \Eprint [0]{\href }%
\providecommand \doibase [0]{https://doi.org/}%
\providecommand \selectlanguage [0]{\@gobble}%
\providecommand \bibinfo  [0]{\@secondoftwo}%
\providecommand \bibfield  [0]{\@secondoftwo}%
\providecommand \translation [1]{[#1]}%
\providecommand \BibitemOpen [0]{}%
\providecommand \bibitemStop [0]{}%
\providecommand \bibitemNoStop [0]{.\EOS\space}%
\providecommand \EOS [0]{\spacefactor3000\relax}%
\providecommand \BibitemShut  [1]{\csname bibitem#1\endcsname}%
\let\auto@bib@innerbib\@empty
\bibitem [{\citenamefont {Polkovnikov}\ \emph {et~al.}(2011)\citenamefont
  {Polkovnikov}, \citenamefont {Sengupta}, \citenamefont {Silva},\ and\
  \citenamefont {Vengalattore}}]{Polkovnikov2011}%
  \BibitemOpen
  \bibfield  {author} {\bibinfo {author} {\bibfnamefont {A.}~\bibnamefont
  {Polkovnikov}}, \bibinfo {author} {\bibfnamefont {K.}~\bibnamefont
  {Sengupta}}, \bibinfo {author} {\bibfnamefont {A.}~\bibnamefont {Silva}},\
  and\ \bibinfo {author} {\bibfnamefont {M.}~\bibnamefont {Vengalattore}},\
  }\bibfield  {title} {\bibinfo {title} {Colloquium: Nonequilibrium dynamics of
  closed interacting quantum systems},\ }\href
  {https://doi.org/10.1103/RevModPhys.83.863} {\bibfield  {journal} {\bibinfo
  {journal} {Rev. Mod. Phys.}\ }\textbf {\bibinfo {volume} {83}},\ \bibinfo
  {pages} {863} (\bibinfo {year} {2011})}\BibitemShut {NoStop}%
\bibitem [{\citenamefont {Mitra}(2018)}]{Mitra2018}%
  \BibitemOpen
  \bibfield  {author} {\bibinfo {author} {\bibfnamefont {A.}~\bibnamefont
  {Mitra}},\ }\bibfield  {title} {\bibinfo {title} {Quantum quench dynamics},\
  }\href {https://doi.org/10.1146/annurev-conmatphys-031016-025451} {\bibfield
  {journal} {\bibinfo  {journal} {Annual Review of Condensed Matter Physics}\
  }\textbf {\bibinfo {volume} {9}},\ \bibinfo {pages} {245} (\bibinfo {year}
  {2018})}\BibitemShut {NoStop}%
\bibitem [{\citenamefont {Castelnovo}\ \emph {et~al.}(2010)\citenamefont
  {Castelnovo}, \citenamefont {Moessner},\ and\ \citenamefont
  {Sondhi}}]{Castelnovo2010}%
  \BibitemOpen
  \bibfield  {author} {\bibinfo {author} {\bibfnamefont {C.}~\bibnamefont
  {Castelnovo}}, \bibinfo {author} {\bibfnamefont {R.}~\bibnamefont
  {Moessner}},\ and\ \bibinfo {author} {\bibfnamefont {S.~L.}\ \bibnamefont
  {Sondhi}},\ }\bibfield  {title} {\bibinfo {title} {Thermal quenches in spin
  ice},\ }\href {https://doi.org/10.1103/PhysRevLett.104.107201} {\bibfield
  {journal} {\bibinfo  {journal} {Phys. Rev. Lett.}\ }\textbf {\bibinfo
  {volume} {104}},\ \bibinfo {pages} {107201} (\bibinfo {year}
  {2010})}\BibitemShut {NoStop}%
\bibitem [{\citenamefont {Mostame}\ \emph {et~al.}(2014)\citenamefont
  {Mostame}, \citenamefont {Castelnovo}, \citenamefont {Moessner},\ and\
  \citenamefont {Sondhi}}]{Mostame2014b}%
  \BibitemOpen
  \bibfield  {author} {\bibinfo {author} {\bibfnamefont {S.}~\bibnamefont
  {Mostame}}, \bibinfo {author} {\bibfnamefont {C.}~\bibnamefont {Castelnovo}},
  \bibinfo {author} {\bibfnamefont {R.}~\bibnamefont {Moessner}},\ and\
  \bibinfo {author} {\bibfnamefont {S.~L.}\ \bibnamefont {Sondhi}},\ }\bibfield
   {title} {\bibinfo {title} {Tunable nonequilibrium dynamics of field quenches
  in spin ice},\ }\href {https://doi.org/10.1073/pnas.1317631111} {\bibfield
  {journal} {\bibinfo  {journal} {Proceedings of the National Academy of
  Sciences}\ }\textbf {\bibinfo {volume} {111}},\ \bibinfo {pages} {640}
  (\bibinfo {year} {2014})}\BibitemShut {NoStop}%
\bibitem [{\citenamefont {Greiner}\ \emph {et~al.}(2002)\citenamefont
  {Greiner}, \citenamefont {Mandel}, \citenamefont {H\"ansch},\ and\
  \citenamefont {Bloch}}]{Greiner2002}%
  \BibitemOpen
  \bibfield  {author} {\bibinfo {author} {\bibfnamefont {M.}~\bibnamefont
  {Greiner}}, \bibinfo {author} {\bibfnamefont {O.}~\bibnamefont {Mandel}},
  \bibinfo {author} {\bibfnamefont {T.~W.}\ \bibnamefont {H\"ansch}},\ and\
  \bibinfo {author} {\bibfnamefont {I.}~\bibnamefont {Bloch}},\ }\bibfield
  {title} {\bibinfo {title} {Collapse and revival of the matter wave field of a
  bose–einstein condensate},\ }\href {https://doi.org/10.1038/nature00968}
  {\bibfield  {journal} {\bibinfo  {journal} {Nature}\ }\textbf {\bibinfo
  {volume} {419}},\ \bibinfo {pages} {51 } (\bibinfo {year}
  {2002})}\BibitemShut {NoStop}%
\bibitem [{\citenamefont {Chen}\ \emph {et~al.}(2011)\citenamefont {Chen},
  \citenamefont {White}, \citenamefont {Borries},\ and\ \citenamefont
  {DeMarco}}]{Chen2011}%
  \BibitemOpen
  \bibfield  {author} {\bibinfo {author} {\bibfnamefont {D.}~\bibnamefont
  {Chen}}, \bibinfo {author} {\bibfnamefont {M.}~\bibnamefont {White}},
  \bibinfo {author} {\bibfnamefont {C.}~\bibnamefont {Borries}},\ and\ \bibinfo
  {author} {\bibfnamefont {B.}~\bibnamefont {DeMarco}},\ }\bibfield  {title}
  {\bibinfo {title} {Quantum quench of an atomic mott insulator},\ }\href
  {https://doi.org/10.1103/PhysRevLett.106.235304} {\bibfield  {journal}
  {\bibinfo  {journal} {Phys. Rev. Lett.}\ }\textbf {\bibinfo {volume} {106}},\
  \bibinfo {pages} {235304} (\bibinfo {year} {2011})}\BibitemShut {NoStop}%
\bibitem [{\citenamefont {\"Olschl\"ager}\ \emph {et~al.}(2011)\citenamefont
  {\"Olschl\"ager}, \citenamefont {Wirth},\ and\ \citenamefont
  {Hemmerich}}]{Matthias2011}%
  \BibitemOpen
  \bibfield  {author} {\bibinfo {author} {\bibfnamefont {M.}~\bibnamefont
  {\"Olschl\"ager}}, \bibinfo {author} {\bibfnamefont {G.}~\bibnamefont
  {Wirth}},\ and\ \bibinfo {author} {\bibfnamefont {A.}~\bibnamefont
  {Hemmerich}},\ }\bibfield  {title} {\bibinfo {title} {Unconventional
  superfluid order in the $f$ band of a bipartite optical square lattice},\
  }\href {https://doi.org/10.1103/PhysRevLett.106.015302} {\bibfield  {journal}
  {\bibinfo  {journal} {Phys. Rev. Lett.}\ }\textbf {\bibinfo {volume} {106}},\
  \bibinfo {pages} {015302} (\bibinfo {year} {2011})}\BibitemShut {NoStop}%
\bibitem [{\citenamefont {Nicklas}\ \emph {et~al.}(2015)\citenamefont
  {Nicklas}, \citenamefont {Karl}, \citenamefont {H\"ofer}, \citenamefont
  {Johnson}, \citenamefont {Muessel}, \citenamefont {Strobel}, \citenamefont
  {Tomkovi\ifmmode~\check{c}\else \v{c}\fi{}}, \citenamefont {Gasenzer},\ and\
  \citenamefont {Oberthaler}}]{Nicklas2015}%
  \BibitemOpen
  \bibfield  {author} {\bibinfo {author} {\bibfnamefont {E.}~\bibnamefont
  {Nicklas}}, \bibinfo {author} {\bibfnamefont {M.}~\bibnamefont {Karl}},
  \bibinfo {author} {\bibfnamefont {M.}~\bibnamefont {H\"ofer}}, \bibinfo
  {author} {\bibfnamefont {A.}~\bibnamefont {Johnson}}, \bibinfo {author}
  {\bibfnamefont {W.}~\bibnamefont {Muessel}}, \bibinfo {author} {\bibfnamefont
  {H.}~\bibnamefont {Strobel}}, \bibinfo {author} {\bibfnamefont
  {J.}~\bibnamefont {Tomkovi\ifmmode~\check{c}\else \v{c}\fi{}}}, \bibinfo
  {author} {\bibfnamefont {T.}~\bibnamefont {Gasenzer}},\ and\ \bibinfo
  {author} {\bibfnamefont {M.~K.}\ \bibnamefont {Oberthaler}},\ }\bibfield
  {title} {\bibinfo {title} {Observation of scaling in the dynamics of a
  strongly quenched quantum gas},\ }\href
  {https://doi.org/10.1103/PhysRevLett.115.245301} {\bibfield  {journal}
  {\bibinfo  {journal} {Phys. Rev. Lett.}\ }\textbf {\bibinfo {volume} {115}},\
  \bibinfo {pages} {245301} (\bibinfo {year} {2015})}\BibitemShut {NoStop}%
\bibitem [{\citenamefont {Pr\"ufer}\ \emph {et~al.}(2018)\citenamefont
  {Pr\"ufer}, \citenamefont {Kunkel}, \citenamefont {Strobel}, \citenamefont
  {Lannig}, \citenamefont {Linnemann}, \citenamefont {Schmied}, \citenamefont
  {Berges}, \citenamefont {Gasenzer},\ and\ \citenamefont
  {Oberthaler}}]{Prufer2018}%
  \BibitemOpen
  \bibfield  {author} {\bibinfo {author} {\bibfnamefont {M.}~\bibnamefont
  {Pr\"ufer}}, \bibinfo {author} {\bibfnamefont {P.}~\bibnamefont {Kunkel}},
  \bibinfo {author} {\bibfnamefont {H.}~\bibnamefont {Strobel}}, \bibinfo
  {author} {\bibfnamefont {S.}~\bibnamefont {Lannig}}, \bibinfo {author}
  {\bibfnamefont {D.}~\bibnamefont {Linnemann}}, \bibinfo {author}
  {\bibfnamefont {C.~M.}\ \bibnamefont {Schmied}}, \bibinfo {author}
  {\bibfnamefont {J.}~\bibnamefont {Berges}}, \bibinfo {author} {\bibfnamefont
  {T.}~\bibnamefont {Gasenzer}},\ and\ \bibinfo {author} {\bibfnamefont
  {M.~K.}\ \bibnamefont {Oberthaler}},\ }\bibfield  {title} {\bibinfo {title}
  {Observation of universal dynamics in a spinor {B}ose gas far from
  equilibrium},\ }\href {https://doi.org/10.1038/s41586-018-0659-0} {\bibfield
  {journal} {\bibinfo  {journal} {Nature}\ }\textbf {\bibinfo {volume} {563}},\
  \bibinfo {pages} {217} (\bibinfo {year} {2018})}\BibitemShut {NoStop}%
\bibitem [{\citenamefont {Paulsen}\ \emph {et~al.}(2014)\citenamefont
  {Paulsen}, \citenamefont {Jackson}, \citenamefont {Lhotel}, \citenamefont
  {Canals}, \citenamefont {Prabhakaran}, \citenamefont {Matsuhira},
  \citenamefont {Giblin},\ and\ \citenamefont {Bramwell}}]{Paulsen2014}%
  \BibitemOpen
  \bibfield  {author} {\bibinfo {author} {\bibfnamefont {C.}~\bibnamefont
  {Paulsen}}, \bibinfo {author} {\bibfnamefont {M.~J.}\ \bibnamefont
  {Jackson}}, \bibinfo {author} {\bibfnamefont {E.}~\bibnamefont {Lhotel}},
  \bibinfo {author} {\bibfnamefont {B.}~\bibnamefont {Canals}}, \bibinfo
  {author} {\bibfnamefont {D.}~\bibnamefont {Prabhakaran}}, \bibinfo {author}
  {\bibfnamefont {K.}~\bibnamefont {Matsuhira}}, \bibinfo {author}
  {\bibfnamefont {S.~R.}\ \bibnamefont {Giblin}},\ and\ \bibinfo {author}
  {\bibfnamefont {S.~T.}\ \bibnamefont {Bramwell}},\ }\bibfield  {title}
  {\bibinfo {title} {Far-from-equilibrium monopole dynamics in spin ice},\
  }\href {https://doi.org/10.1038/nphys2847} {\bibfield  {journal} {\bibinfo
  {journal} {Nature Physics}\ }\textbf {\bibinfo {volume} {10}},\ \bibinfo
  {pages} {135} (\bibinfo {year} {2014})}\BibitemShut {NoStop}%
\bibitem [{\citenamefont {Harris}\ \emph {et~al.}(1997)\citenamefont {Harris},
  \citenamefont {Bramwell}, \citenamefont {McMorrow}, \citenamefont {Zeiske},\
  and\ \citenamefont {Godfrey}}]{Harris1997}%
  \BibitemOpen
  \bibfield  {author} {\bibinfo {author} {\bibfnamefont {M.~J.}\ \bibnamefont
  {Harris}}, \bibinfo {author} {\bibfnamefont {S.~T.}\ \bibnamefont
  {Bramwell}}, \bibinfo {author} {\bibfnamefont {D.~F.}\ \bibnamefont
  {McMorrow}}, \bibinfo {author} {\bibfnamefont {T.}~\bibnamefont {Zeiske}},\
  and\ \bibinfo {author} {\bibfnamefont {K.~W.}\ \bibnamefont {Godfrey}},\
  }\bibfield  {title} {\bibinfo {title} {Geometrical frustration in the
  ferromagnetic pyrochlore {H}o$_{2}${T}i$_{2}${O}$_{7}$},\ }\href
  {https://doi.org/10.1103/PhysRevLett.79.2554} {\bibfield  {journal} {\bibinfo
   {journal} {Phys. Rev. Lett.}\ }\textbf {\bibinfo {volume} {79}},\ \bibinfo
  {pages} {2554} (\bibinfo {year} {1997})}\BibitemShut {NoStop}%
\bibitem [{\citenamefont {Gardner}\ \emph {et~al.}(2010)\citenamefont
  {Gardner}, \citenamefont {Gingras},\ and\ \citenamefont
  {Greedan}}]{Gardiner2010}%
  \BibitemOpen
  \bibfield  {author} {\bibinfo {author} {\bibfnamefont {J.~S.}\ \bibnamefont
  {Gardner}}, \bibinfo {author} {\bibfnamefont {M.~J.~P.}\ \bibnamefont
  {Gingras}},\ and\ \bibinfo {author} {\bibfnamefont {J.~E.}\ \bibnamefont
  {Greedan}},\ }\bibfield  {title} {\bibinfo {title} {Magnetic pyrochlore
  oxides},\ }\href {https://doi.org/10.1103/RevModPhys.82.53} {\bibfield
  {journal} {\bibinfo  {journal} {Rev. Mod. Phys.}\ }\textbf {\bibinfo {volume}
  {82}},\ \bibinfo {pages} {53} (\bibinfo {year} {2010})}\BibitemShut {NoStop}%
\bibitem [{\citenamefont {Isakov}\ \emph {et~al.}(2005)\citenamefont {Isakov},
  \citenamefont {Moessner},\ and\ \citenamefont {Sondhi}}]{Isakov2005}%
  \BibitemOpen
  \bibfield  {author} {\bibinfo {author} {\bibfnamefont {S.~V.}\ \bibnamefont
  {Isakov}}, \bibinfo {author} {\bibfnamefont {R.}~\bibnamefont {Moessner}},\
  and\ \bibinfo {author} {\bibfnamefont {S.~L.}\ \bibnamefont {Sondhi}},\
  }\bibfield  {title} {\bibinfo {title} {Why spin ice obeys the ice rules},\
  }\href {https://doi.org/10.1103/PhysRevLett.95.217201} {\bibfield  {journal}
  {\bibinfo  {journal} {Phys. Rev. Lett.}\ }\textbf {\bibinfo {volume} {95}},\
  \bibinfo {pages} {217201} (\bibinfo {year} {2005})}\BibitemShut {NoStop}%
\bibitem [{\citenamefont {Castelnovo}\ \emph {et~al.}(2008)\citenamefont
  {Castelnovo}, \citenamefont {Moessner},\ and\ \citenamefont
  {Sondhi}}]{Castelnovo2008}%
  \BibitemOpen
  \bibfield  {author} {\bibinfo {author} {\bibfnamefont {C.}~\bibnamefont
  {Castelnovo}}, \bibinfo {author} {\bibfnamefont {R.}~\bibnamefont
  {Moessner}},\ and\ \bibinfo {author} {\bibfnamefont {S.~L.}\ \bibnamefont
  {Sondhi}},\ }\bibfield  {title} {\bibinfo {title} {Magnetic monopoles in spin
  ice},\ }\href {https://doi.org/10.1038/nature06433} {\bibfield  {journal}
  {\bibinfo  {journal} {Nature}\ }\textbf {\bibinfo {volume} {451}},\ \bibinfo
  {pages} {42} (\bibinfo {year} {2008})}\BibitemShut {NoStop}%
\bibitem [{\citenamefont {Ramirez}\ \emph {et~al.}(1999)\citenamefont
  {Ramirez}, \citenamefont {Hayashi}, \citenamefont {Cava}, \citenamefont
  {Siddharthan},\ and\ \citenamefont {Shastry}}]{Ramirez1999}%
  \BibitemOpen
  \bibfield  {author} {\bibinfo {author} {\bibfnamefont {A.~P.}\ \bibnamefont
  {Ramirez}}, \bibinfo {author} {\bibfnamefont {A.}~\bibnamefont {Hayashi}},
  \bibinfo {author} {\bibfnamefont {R.~J.}\ \bibnamefont {Cava}}, \bibinfo
  {author} {\bibfnamefont {R.}~\bibnamefont {Siddharthan}},\ and\ \bibinfo
  {author} {\bibfnamefont {B.~S.}\ \bibnamefont {Shastry}},\ }\bibfield
  {title} {\bibinfo {title} {Zero-point entropy in ‘spin ice’},\ }\href
  {https://doi.org/10.1038/20619} {\bibfield  {journal} {\bibinfo  {journal}
  {Nature}\ }\textbf {\bibinfo {volume} {399}},\ \bibinfo {pages} {333}
  (\bibinfo {year} {1999})}\BibitemShut {NoStop}%
\bibitem [{\citenamefont {Rosenkranz}\ \emph {et~al.}(2000)\citenamefont
  {Rosenkranz}, \citenamefont {Ramirez}, \citenamefont {Hayashi}, \citenamefont
  {Cava}, \citenamefont {Siddharthan},\ and\ \citenamefont
  {Shastry}}]{Rosenkranz2000}%
  \BibitemOpen
  \bibfield  {author} {\bibinfo {author} {\bibfnamefont {S.}~\bibnamefont
  {Rosenkranz}}, \bibinfo {author} {\bibfnamefont {A.~P.}\ \bibnamefont
  {Ramirez}}, \bibinfo {author} {\bibfnamefont {A.}~\bibnamefont {Hayashi}},
  \bibinfo {author} {\bibfnamefont {R.~J.}\ \bibnamefont {Cava}}, \bibinfo
  {author} {\bibfnamefont {R.}~\bibnamefont {Siddharthan}},\ and\ \bibinfo
  {author} {\bibfnamefont {B.~S.}\ \bibnamefont {Shastry}},\ }\bibfield
  {title} {\bibinfo {title} {Crystal-field interaction in the pyrochlore magnet
  {${\mathrm{Ho}}_{2}{\mathrm{Ti}}_{2}{\mathrm{O}}_{7}$}},\ }\href
  {https://doi.org/10.1063/1.372565} {\bibfield  {journal} {\bibinfo  {journal}
  {Journal of Applied Physics}\ }\textbf {\bibinfo {volume} {87}},\ \bibinfo
  {pages} {5914} (\bibinfo {year} {2000})}\BibitemShut {NoStop}%
\bibitem [{\citenamefont {Bramwell}\ \emph {et~al.}(2001)\citenamefont
  {Bramwell}, \citenamefont {Harris}, \citenamefont {den Hertog}, \citenamefont
  {Gingras}, \citenamefont {Gardner}, \citenamefont {McMorrow}, \citenamefont
  {Wildes}, \citenamefont {Cornelius}, \citenamefont {Champion}, \citenamefont
  {Melko},\ and\ \citenamefont {Fennell}}]{Bramwell2001b}%
  \BibitemOpen
  \bibfield  {author} {\bibinfo {author} {\bibfnamefont {S.~T.}\ \bibnamefont
  {Bramwell}}, \bibinfo {author} {\bibfnamefont {M.~J.}\ \bibnamefont
  {Harris}}, \bibinfo {author} {\bibfnamefont {B.~C.}\ \bibnamefont {den
  Hertog}}, \bibinfo {author} {\bibfnamefont {M.~J.~P.}\ \bibnamefont
  {Gingras}}, \bibinfo {author} {\bibfnamefont {J.~S.}\ \bibnamefont
  {Gardner}}, \bibinfo {author} {\bibfnamefont {D.~F.}\ \bibnamefont
  {McMorrow}}, \bibinfo {author} {\bibfnamefont {A.~R.}\ \bibnamefont
  {Wildes}}, \bibinfo {author} {\bibfnamefont {A.~L.}\ \bibnamefont
  {Cornelius}}, \bibinfo {author} {\bibfnamefont {J.~D.~M.}\ \bibnamefont
  {Champion}}, \bibinfo {author} {\bibfnamefont {R.~G.}\ \bibnamefont
  {Melko}},\ and\ \bibinfo {author} {\bibfnamefont {T.}~\bibnamefont
  {Fennell}},\ }\bibfield  {title} {\bibinfo {title} {Spin correlations in
  {H}o$_{2}${T}i$_{2}${O}$_{7}$: A dipolar spin ice system},\ }\href
  {https://doi.org/10.1103/PhysRevLett.87.047205} {\bibfield  {journal}
  {\bibinfo  {journal} {Phys. Rev. Lett.}\ }\textbf {\bibinfo {volume} {87}},\
  \bibinfo {pages} {047205} (\bibinfo {year} {2001})}\BibitemShut {NoStop}%
\bibitem [{\citenamefont {Rau}\ and\ \citenamefont {Gingras}(2015)}]{Rau2015}%
  \BibitemOpen
  \bibfield  {author} {\bibinfo {author} {\bibfnamefont {J.~G.}\ \bibnamefont
  {Rau}}\ and\ \bibinfo {author} {\bibfnamefont {M.~J.~P.}\ \bibnamefont
  {Gingras}},\ }\bibfield  {title} {\bibinfo {title} {Magnitude of quantum
  effects in classical spin ices},\ }\href
  {https://doi.org/10.1103/PhysRevB.92.144417} {\bibfield  {journal} {\bibinfo
  {journal} {Phys. Rev. B}\ }\textbf {\bibinfo {volume} {92}},\ \bibinfo
  {pages} {144417} (\bibinfo {year} {2015})}\BibitemShut {NoStop}%
\bibitem [{\citenamefont {Gingras}\ and\ \citenamefont
  {McClarty}(2014)}]{Gingras2014}%
  \BibitemOpen
  \bibfield  {author} {\bibinfo {author} {\bibfnamefont {M.~J.~P.}\
  \bibnamefont {Gingras}}\ and\ \bibinfo {author} {\bibfnamefont {P.~A.}\
  \bibnamefont {McClarty}},\ }\bibfield  {title} {\bibinfo {title} {Quantum
  spin ice: a search for gapless quantum spin liquids in pyrochlore magnets},\
  }\href {https://doi.org/10.1088/0034-4885/77/5/056501} {\bibfield  {journal}
  {\bibinfo  {journal} {Reports on Progress in Physics}\ }\textbf {\bibinfo
  {volume} {77}},\ \bibinfo {pages} {056501} (\bibinfo {year}
  {2014})}\BibitemShut {NoStop}%
\bibitem [{\citenamefont {Giblin}\ \emph {et~al.}(2018)\citenamefont {Giblin},
  \citenamefont {Twengstr\"om}, \citenamefont {Bovo}, \citenamefont {Ruminy},
  \citenamefont {Bartkowiak}, \citenamefont {Manuel}, \citenamefont {Andresen},
  \citenamefont {Prabhakaran}, \citenamefont {Balakrishnan}, \citenamefont
  {Pomjakushina}, \citenamefont {Paulsen}, \citenamefont {Lhotel},
  \citenamefont {Keller}, \citenamefont {Frontzek}, \citenamefont {Capelli},
  \citenamefont {Zaharko}, \citenamefont {McClarty}, \citenamefont {Bramwell},
  \citenamefont {Henelius},\ and\ \citenamefont {Fennell}}]{Giblin2018}%
  \BibitemOpen
  \bibfield  {author} {\bibinfo {author} {\bibfnamefont {S.~R.}\ \bibnamefont
  {Giblin}}, \bibinfo {author} {\bibfnamefont {M.}~\bibnamefont
  {Twengstr\"om}}, \bibinfo {author} {\bibfnamefont {L.}~\bibnamefont {Bovo}},
  \bibinfo {author} {\bibfnamefont {M.}~\bibnamefont {Ruminy}}, \bibinfo
  {author} {\bibfnamefont {M.}~\bibnamefont {Bartkowiak}}, \bibinfo {author}
  {\bibfnamefont {P.}~\bibnamefont {Manuel}}, \bibinfo {author} {\bibfnamefont
  {J.~C.}\ \bibnamefont {Andresen}}, \bibinfo {author} {\bibfnamefont
  {D.}~\bibnamefont {Prabhakaran}}, \bibinfo {author} {\bibfnamefont
  {G.}~\bibnamefont {Balakrishnan}}, \bibinfo {author} {\bibfnamefont
  {E.}~\bibnamefont {Pomjakushina}}, \bibinfo {author} {\bibfnamefont
  {C.}~\bibnamefont {Paulsen}}, \bibinfo {author} {\bibfnamefont
  {E.}~\bibnamefont {Lhotel}}, \bibinfo {author} {\bibfnamefont
  {L.}~\bibnamefont {Keller}}, \bibinfo {author} {\bibfnamefont
  {M.}~\bibnamefont {Frontzek}}, \bibinfo {author} {\bibfnamefont {S.~C.}\
  \bibnamefont {Capelli}}, \bibinfo {author} {\bibfnamefont {O.}~\bibnamefont
  {Zaharko}}, \bibinfo {author} {\bibfnamefont {P.~A.}\ \bibnamefont
  {McClarty}}, \bibinfo {author} {\bibfnamefont {S.~T.}\ \bibnamefont
  {Bramwell}}, \bibinfo {author} {\bibfnamefont {P.}~\bibnamefont {Henelius}},\
  and\ \bibinfo {author} {\bibfnamefont {T.}~\bibnamefont {Fennell}},\
  }\bibfield  {title} {\bibinfo {title} {Pauling entropy, metastability, and
  equilibrium in {D}y$_{2}${T}i$_{2}${O}$_{7}$ spin ice},\ }\href
  {https://doi.org/10.1103/PhysRevLett.121.067202} {\bibfield  {journal}
  {\bibinfo  {journal} {Phys. Rev. Lett.}\ }\textbf {\bibinfo {volume} {121}},\
  \bibinfo {pages} {067202} (\bibinfo {year} {2018})}\BibitemShut {NoStop}%
\bibitem [{\citenamefont {Fennell}\ \emph {et~al.}(2009)\citenamefont
  {Fennell}, \citenamefont {Deen}, \citenamefont {Wildes}, \citenamefont
  {Schmalzl}, \citenamefont {Prabhakaran}, \citenamefont {Boothroyd},
  \citenamefont {Aldus}, \citenamefont {McMorrow},\ and\ \citenamefont
  {Bramwell}}]{Fennell2009}%
  \BibitemOpen
  \bibfield  {author} {\bibinfo {author} {\bibfnamefont {T.}~\bibnamefont
  {Fennell}}, \bibinfo {author} {\bibfnamefont {P.~P.}\ \bibnamefont {Deen}},
  \bibinfo {author} {\bibfnamefont {A.~R.}\ \bibnamefont {Wildes}}, \bibinfo
  {author} {\bibfnamefont {K.}~\bibnamefont {Schmalzl}}, \bibinfo {author}
  {\bibfnamefont {D.}~\bibnamefont {Prabhakaran}}, \bibinfo {author}
  {\bibfnamefont {A.~T.}\ \bibnamefont {Boothroyd}}, \bibinfo {author}
  {\bibfnamefont {R.~J.}\ \bibnamefont {Aldus}}, \bibinfo {author}
  {\bibfnamefont {D.~F.}\ \bibnamefont {McMorrow}},\ and\ \bibinfo {author}
  {\bibfnamefont {S.~T.}\ \bibnamefont {Bramwell}},\ }\bibfield  {title}
  {\bibinfo {title} {Magnetic {C}oulomb phase in the spin ice
  {H}o$_{2}${T}i$_{2}${O}$_{7}$},\ }\href
  {https://doi.org/10.1126/science.1177582} {\bibfield  {journal} {\bibinfo
  {journal} {Science}\ }\textbf {\bibinfo {volume} {326}},\ \bibinfo {pages}
  {415} (\bibinfo {year} {2009})}\BibitemShut {NoStop}%
\bibitem [{\citenamefont {Henley}(2010)}]{Henley2010}%
  \BibitemOpen
  \bibfield  {author} {\bibinfo {author} {\bibfnamefont {C.~L.}\ \bibnamefont
  {Henley}},\ }\bibfield  {title} {\bibinfo {title} {The
  {\textquotedblleft}{C}oulomb phase{\textquotedblright} in frustrated
  systems},\ }\href {https://doi.org/10.1146/annurev-conmatphys-070909-104138}
  {\bibfield  {journal} {\bibinfo  {journal} {Annual Review of Condensed Matter
  Physics}\ }\textbf {\bibinfo {volume} {1}},\ \bibinfo {pages} {179} (\bibinfo
  {year} {2010})}\BibitemShut {NoStop}%
\bibitem [{\citenamefont {den Hertog}\ and\ \citenamefont
  {Gingras}(2000)}]{denHertog2000}%
  \BibitemOpen
  \bibfield  {author} {\bibinfo {author} {\bibfnamefont {B.~C.}\ \bibnamefont
  {den Hertog}}\ and\ \bibinfo {author} {\bibfnamefont {M.~J.~P.}\ \bibnamefont
  {Gingras}},\ }\bibfield  {title} {\bibinfo {title} {Dipolar interactions and
  origin of spin ice in {I}sing pyrochlore magnets},\ }\href
  {https://doi.org/10.1103/PhysRevLett.84.3430} {\bibfield  {journal} {\bibinfo
   {journal} {Phys. Rev. Lett.}\ }\textbf {\bibinfo {volume} {84}},\ \bibinfo
  {pages} {3430} (\bibinfo {year} {2000})}\BibitemShut {NoStop}%
\bibitem [{\citenamefont {Castelnovo}\ \emph {et~al.}(2012)\citenamefont
  {Castelnovo}, \citenamefont {Moessner},\ and\ \citenamefont
  {Sondhi}}]{Castelnovo2012}%
  \BibitemOpen
  \bibfield  {author} {\bibinfo {author} {\bibfnamefont {C.}~\bibnamefont
  {Castelnovo}}, \bibinfo {author} {\bibfnamefont {R.}~\bibnamefont
  {Moessner}},\ and\ \bibinfo {author} {\bibfnamefont {S.}~\bibnamefont
  {Sondhi}},\ }\bibfield  {title} {\bibinfo {title} {Spin ice,
  fractionalization, and topological order},\ }\href
  {https://doi.org/10.1146/annurev-conmatphys-020911-125058} {\bibfield
  {journal} {\bibinfo  {journal} {Annual Review of Condensed Matter Physics}\
  }\textbf {\bibinfo {volume} {3}},\ \bibinfo {pages} {35} (\bibinfo {year}
  {2012})}\BibitemShut {NoStop}%
\bibitem [{\citenamefont {Kadowaki}\ \emph {et~al.}(2009)\citenamefont
  {Kadowaki}, \citenamefont {Doi}, \citenamefont {Aoki}, \citenamefont
  {Tabata}, \citenamefont {J.~Sato}, \citenamefont {W.~Lynn}, \citenamefont
  {Matsuhira},\ and\ \citenamefont {Hiroi}}]{Kadowaki2009}%
  \BibitemOpen
  \bibfield  {author} {\bibinfo {author} {\bibfnamefont {H.}~\bibnamefont
  {Kadowaki}}, \bibinfo {author} {\bibfnamefont {N.}~\bibnamefont {Doi}},
  \bibinfo {author} {\bibfnamefont {Y.}~\bibnamefont {Aoki}}, \bibinfo {author}
  {\bibfnamefont {Y.}~\bibnamefont {Tabata}}, \bibinfo {author} {\bibfnamefont
  {T.}~\bibnamefont {J.~Sato}}, \bibinfo {author} {\bibfnamefont
  {J.}~\bibnamefont {W.~Lynn}}, \bibinfo {author} {\bibfnamefont
  {K.}~\bibnamefont {Matsuhira}},\ and\ \bibinfo {author} {\bibfnamefont
  {Z.}~\bibnamefont {Hiroi}},\ }\bibfield  {title} {\bibinfo {title}
  {Observation of magnetic monopoles in spin ice},\ }\href
  {https://doi.org/10.1143/JPSJ.78.103706} {\bibfield  {journal} {\bibinfo
  {journal} {Journal of the Physical Society of Japan}\ }\textbf {\bibinfo
  {volume} {78}},\ \bibinfo {pages} {103706} (\bibinfo {year}
  {2009})}\BibitemShut {NoStop}%
\bibitem [{\citenamefont {Giblin}\ \emph {et~al.}(2011)\citenamefont {Giblin},
  \citenamefont {Bramwell}, \citenamefont {Holdsworth}, \citenamefont
  {Prabhakaran},\ and\ \citenamefont {Terry}}]{Giblin2011}%
  \BibitemOpen
  \bibfield  {author} {\bibinfo {author} {\bibfnamefont {S.~R.}\ \bibnamefont
  {Giblin}}, \bibinfo {author} {\bibfnamefont {S.~T.}\ \bibnamefont
  {Bramwell}}, \bibinfo {author} {\bibfnamefont {P.~C.~W.}\ \bibnamefont
  {Holdsworth}}, \bibinfo {author} {\bibfnamefont {D.}~\bibnamefont
  {Prabhakaran}},\ and\ \bibinfo {author} {\bibfnamefont {I.}~\bibnamefont
  {Terry}},\ }\bibfield  {title} {\bibinfo {title} {Creation and measurement of
  long-lived magnetic monopole currents in spin ice},\ }\href
  {https://doi.org/10.1038/nphys1896} {\bibfield  {journal} {\bibinfo
  {journal} {Nature Physics}\ }\textbf {\bibinfo {volume} {7}},\ \bibinfo
  {pages} {252} (\bibinfo {year} {2011})}\BibitemShut {NoStop}%
\bibitem [{\citenamefont {Fennell}\ \emph {et~al.}(2005)\citenamefont
  {Fennell}, \citenamefont {Petrenko}, \citenamefont {F\aa{}k}, \citenamefont
  {Gardner}, \citenamefont {Bramwell},\ and\ \citenamefont
  {Ouladdiaf}}]{Fennell2005}%
  \BibitemOpen
  \bibfield  {author} {\bibinfo {author} {\bibfnamefont {T.}~\bibnamefont
  {Fennell}}, \bibinfo {author} {\bibfnamefont {O.~A.}\ \bibnamefont
  {Petrenko}}, \bibinfo {author} {\bibfnamefont {B.}~\bibnamefont {F\aa{}k}},
  \bibinfo {author} {\bibfnamefont {J.~S.}\ \bibnamefont {Gardner}}, \bibinfo
  {author} {\bibfnamefont {S.~T.}\ \bibnamefont {Bramwell}},\ and\ \bibinfo
  {author} {\bibfnamefont {B.}~\bibnamefont {Ouladdiaf}},\ }\bibfield  {title}
  {\bibinfo {title} {Neutron scattering studies of the spin ices
  {${\mathrm{Ho}}_{2}{\mathrm{Ti}}_{2}{\mathrm{O}}_{7}$ and
  ${\mathrm{Dy}}_{2}{\mathrm{Ti}}_{2}{\mathrm{O}}_{7}$} in applied magnetic
  field},\ }\href {https://doi.org/10.1103/PhysRevB.72.224411} {\bibfield
  {journal} {\bibinfo  {journal} {Phys. Rev. B}\ }\textbf {\bibinfo {volume}
  {72}},\ \bibinfo {pages} {224411} (\bibinfo {year} {2005})}\BibitemShut
  {NoStop}%
\bibitem [{\citenamefont {Jaubert}\ \emph {et~al.}(2008)\citenamefont
  {Jaubert}, \citenamefont {Chalker}, \citenamefont {Holdsworth},\ and\
  \citenamefont {Moessner}}]{Jaubert2008}%
  \BibitemOpen
  \bibfield  {author} {\bibinfo {author} {\bibfnamefont {L.~D.~C.}\
  \bibnamefont {Jaubert}}, \bibinfo {author} {\bibfnamefont {J.~T.}\
  \bibnamefont {Chalker}}, \bibinfo {author} {\bibfnamefont {P.~C.~W.}\
  \bibnamefont {Holdsworth}},\ and\ \bibinfo {author} {\bibfnamefont
  {R.}~\bibnamefont {Moessner}},\ }\bibfield  {title} {\bibinfo {title}
  {Three-dimensional {K}asteleyn transition: Spin ice in a [100] field},\
  }\href {https://doi.org/10.1103/PhysRevLett.100.067207} {\bibfield  {journal}
  {\bibinfo  {journal} {Phys. Rev. Lett.}\ }\textbf {\bibinfo {volume} {100}},\
  \bibinfo {pages} {067207} (\bibinfo {year} {2008})}\BibitemShut {NoStop}%
\bibitem [{\citenamefont {Morris}\ \emph {et~al.}(2009)\citenamefont {Morris},
  \citenamefont {Tennant}, \citenamefont {Grigera}, \citenamefont {Klemke},
  \citenamefont {Castelnovo}, \citenamefont {Moessner}, \citenamefont
  {Czternasty}, \citenamefont {Meissner}, \citenamefont {Rule}, \citenamefont
  {Hoffmann}, \citenamefont {Kiefer}, \citenamefont {Gerischer}, \citenamefont
  {Slobinsky},\ and\ \citenamefont {Perry}}]{Morris2009}%
  \BibitemOpen
  \bibfield  {author} {\bibinfo {author} {\bibfnamefont {D.~J.~P.}\
  \bibnamefont {Morris}}, \bibinfo {author} {\bibfnamefont {D.~A.}\
  \bibnamefont {Tennant}}, \bibinfo {author} {\bibfnamefont {S.~A.}\
  \bibnamefont {Grigera}}, \bibinfo {author} {\bibfnamefont {B.}~\bibnamefont
  {Klemke}}, \bibinfo {author} {\bibfnamefont {C.}~\bibnamefont {Castelnovo}},
  \bibinfo {author} {\bibfnamefont {R.}~\bibnamefont {Moessner}}, \bibinfo
  {author} {\bibfnamefont {C.}~\bibnamefont {Czternasty}}, \bibinfo {author}
  {\bibfnamefont {M.}~\bibnamefont {Meissner}}, \bibinfo {author}
  {\bibfnamefont {K.~C.}\ \bibnamefont {Rule}}, \bibinfo {author}
  {\bibfnamefont {J.-U.}\ \bibnamefont {Hoffmann}}, \bibinfo {author}
  {\bibfnamefont {K.}~\bibnamefont {Kiefer}}, \bibinfo {author} {\bibfnamefont
  {S.}~\bibnamefont {Gerischer}}, \bibinfo {author} {\bibfnamefont
  {D.}~\bibnamefont {Slobinsky}},\ and\ \bibinfo {author} {\bibfnamefont
  {R.~S.}\ \bibnamefont {Perry}},\ }\bibfield  {title} {\bibinfo {title} {Dirac
  strings and magnetic monopoles in the spin ice
  {D}y$_{2}${T}i$_{2}${O}$_{7}$},\ }\href
  {https://doi.org/10.1126/science.1178868} {\bibfield  {journal} {\bibinfo
  {journal} {Science}\ }\textbf {\bibinfo {volume} {326}},\ \bibinfo {pages}
  {411} (\bibinfo {year} {2009})}\BibitemShut {NoStop}%
\bibitem [{\citenamefont {Kasteleyn}(1963)}]{Kasteleyn1963}%
  \BibitemOpen
  \bibfield  {author} {\bibinfo {author} {\bibfnamefont {P.~W.}\ \bibnamefont
  {Kasteleyn}},\ }\bibfield  {title} {\bibinfo {title} {Dimer statistics and
  phase transitions},\ }\href {https://doi.org/10.1063/1.1703953} {\bibfield
  {journal} {\bibinfo  {journal} {Journal of Mathematical Physics}\ }\textbf
  {\bibinfo {volume} {4}},\ \bibinfo {pages} {287} (\bibinfo {year}
  {1963})}\BibitemShut {NoStop}%
\bibitem [{\citenamefont {Powell}\ and\ \citenamefont
  {Chalker}(2008)}]{Powell2008}%
  \BibitemOpen
  \bibfield  {author} {\bibinfo {author} {\bibfnamefont {S.}~\bibnamefont
  {Powell}}\ and\ \bibinfo {author} {\bibfnamefont {J.~T.}\ \bibnamefont
  {Chalker}},\ }\bibfield  {title} {\bibinfo {title} {Classical to quantum
  mappings for geometrically frustrated systems: Spin-ice in a [100] field},\
  }\href {https://doi.org/10.1103/PhysRevB.78.024422} {\bibfield  {journal}
  {\bibinfo  {journal} {Phys. Rev. B}\ }\textbf {\bibinfo {volume} {78}},\
  \bibinfo {pages} {024422} (\bibinfo {year} {2008})}\BibitemShut {NoStop}%
\bibitem [{\citenamefont {Powell}(2013)}]{Powell2013}%
  \BibitemOpen
  \bibfield  {author} {\bibinfo {author} {\bibfnamefont {S.}~\bibnamefont
  {Powell}},\ }\bibfield  {title} {\bibinfo {title} {Confinement of monopoles
  and scaling theory near unconventional critical points},\ }\href
  {https://doi.org/10.1103/PhysRevB.87.064414} {\bibfield  {journal} {\bibinfo
  {journal} {Phys. Rev. B}\ }\textbf {\bibinfo {volume} {87}},\ \bibinfo
  {pages} {064414} (\bibinfo {year} {2013})}\BibitemShut {NoStop}%
\bibitem [{\citenamefont {Bramwell}\ and\ \citenamefont
  {Gingras}(2001)}]{Bramwell2001}%
  \BibitemOpen
  \bibfield  {author} {\bibinfo {author} {\bibfnamefont {S.~T.}\ \bibnamefont
  {Bramwell}}\ and\ \bibinfo {author} {\bibfnamefont {M.~J.~P.}\ \bibnamefont
  {Gingras}},\ }\bibfield  {title} {\bibinfo {title} {Spin ice state in
  frustrated magnetic pyrochlore materials},\ }\href
  {https://doi.org/10.1126/science.1064761} {\bibfield  {journal} {\bibinfo
  {journal} {Science}\ }\textbf {\bibinfo {volume} {294}},\ \bibinfo {pages}
  {1495} (\bibinfo {year} {2001})}\BibitemShut {NoStop}%
\bibitem [{\citenamefont {Fennell}\ \emph {et~al.}(2004)\citenamefont
  {Fennell}, \citenamefont {Petrenko}, \citenamefont {F\aa{}k}, \citenamefont
  {Bramwell}, \citenamefont {Enjalran}, \citenamefont {Yavors'kii},
  \citenamefont {Gingras}, \citenamefont {Melko},\ and\ \citenamefont
  {Balakrishnan}}]{Fennell2004}%
  \BibitemOpen
  \bibfield  {author} {\bibinfo {author} {\bibfnamefont {T.}~\bibnamefont
  {Fennell}}, \bibinfo {author} {\bibfnamefont {O.~A.}\ \bibnamefont
  {Petrenko}}, \bibinfo {author} {\bibfnamefont {B.}~\bibnamefont {F\aa{}k}},
  \bibinfo {author} {\bibfnamefont {S.~T.}\ \bibnamefont {Bramwell}}, \bibinfo
  {author} {\bibfnamefont {M.}~\bibnamefont {Enjalran}}, \bibinfo {author}
  {\bibfnamefont {T.}~\bibnamefont {Yavors'kii}}, \bibinfo {author}
  {\bibfnamefont {M.~J.~P.}\ \bibnamefont {Gingras}}, \bibinfo {author}
  {\bibfnamefont {R.~G.}\ \bibnamefont {Melko}},\ and\ \bibinfo {author}
  {\bibfnamefont {G.}~\bibnamefont {Balakrishnan}},\ }\bibfield  {title}
  {\bibinfo {title} {Neutron scattering investigation of the spin ice state in
  {${\mathrm{Dy}}_{2}{\mathrm{Ti}}_{2}{\mathrm{O}}_{7}$}},\ }\href
  {https://doi.org/10.1103/PhysRevB.70.134408} {\bibfield  {journal} {\bibinfo
  {journal} {Phys. Rev. B}\ }\textbf {\bibinfo {volume} {70}},\ \bibinfo
  {pages} {134408} (\bibinfo {year} {2004})}\BibitemShut {NoStop}%
\bibitem [{\citenamefont {Yavors'kii}\ \emph {et~al.}(2008)\citenamefont
  {Yavors'kii}, \citenamefont {Fennell}, \citenamefont {Gingras},\ and\
  \citenamefont {Bramwell}}]{Yavorskii2008}%
  \BibitemOpen
  \bibfield  {author} {\bibinfo {author} {\bibfnamefont {T.}~\bibnamefont
  {Yavors'kii}}, \bibinfo {author} {\bibfnamefont {T.}~\bibnamefont {Fennell}},
  \bibinfo {author} {\bibfnamefont {M.~J.~P.}\ \bibnamefont {Gingras}},\ and\
  \bibinfo {author} {\bibfnamefont {S.~T.}\ \bibnamefont {Bramwell}},\
  }\bibfield  {title} {\bibinfo {title}
  {{${\mathrm{Dy}}_{2}{\mathrm{Ti}}_{2}{\mathrm{O}}_{7}$} spin ice: A test case
  for emergent clusters in a frustrated magnet},\ }\href
  {https://doi.org/10.1103/PhysRevLett.101.037204} {\bibfield  {journal}
  {\bibinfo  {journal} {Phys. Rev. Lett.}\ }\textbf {\bibinfo {volume} {101}},\
  \bibinfo {pages} {037204} (\bibinfo {year} {2008})}\BibitemShut {NoStop}%
\bibitem [{\citenamefont {Bovo}\ \emph {et~al.}(2018)\citenamefont {Bovo},
  \citenamefont {Twengstr{\"o}m}, \citenamefont {Petrenko}, \citenamefont
  {Fennell}, \citenamefont {Gingras}, \citenamefont {Bramwell},\ and\
  \citenamefont {Henelius}}]{Bovo2018}%
  \BibitemOpen
  \bibfield  {author} {\bibinfo {author} {\bibfnamefont {L.}~\bibnamefont
  {Bovo}}, \bibinfo {author} {\bibfnamefont {M.}~\bibnamefont
  {Twengstr{\"o}m}}, \bibinfo {author} {\bibfnamefont {O.~A.}\ \bibnamefont
  {Petrenko}}, \bibinfo {author} {\bibfnamefont {T.}~\bibnamefont {Fennell}},
  \bibinfo {author} {\bibfnamefont {M.~J.~P.}\ \bibnamefont {Gingras}},
  \bibinfo {author} {\bibfnamefont {S.~T.}\ \bibnamefont {Bramwell}},\ and\
  \bibinfo {author} {\bibfnamefont {P.}~\bibnamefont {Henelius}},\ }\bibfield
  {title} {\bibinfo {title} {Special temperatures in frustrated ferromagnets},\
  }\href {https://doi.org/10.1038/s41467-018-04297-3} {\bibfield  {journal}
  {\bibinfo  {journal} {Nature Communications}\ }\textbf {\bibinfo {volume}
  {9}},\ \bibinfo {pages} {1999} (\bibinfo {year} {2018})}\BibitemShut
  {NoStop}%
\bibitem [{\citenamefont {Raban}\ \emph {et~al.}(2019)\citenamefont {Raban},
  \citenamefont {Suen}, \citenamefont {Berthier},\ and\ \citenamefont
  {Holdsworth}}]{Raban2019}%
  \BibitemOpen
  \bibfield  {author} {\bibinfo {author} {\bibfnamefont {V.}~\bibnamefont
  {Raban}}, \bibinfo {author} {\bibfnamefont {C.~T.}\ \bibnamefont {Suen}},
  \bibinfo {author} {\bibfnamefont {L.}~\bibnamefont {Berthier}},\ and\
  \bibinfo {author} {\bibfnamefont {P.~C.~W.}\ \bibnamefont {Holdsworth}},\
  }\bibfield  {title} {\bibinfo {title} {Multiple symmetry sustaining phase
  transitions in spin ice},\ }\href
  {https://doi.org/10.1103/PhysRevB.99.224425} {\bibfield  {journal} {\bibinfo
  {journal} {Phys. Rev. B}\ }\textbf {\bibinfo {volume} {99}},\ \bibinfo
  {pages} {224425} (\bibinfo {year} {2019})}\BibitemShut {NoStop}%
\bibitem [{\citenamefont {de~Leeuw}\ \emph {et~al.}(1980)\citenamefont
  {de~Leeuw}, \citenamefont {Perram},\ and\ \citenamefont {Smith}}]{Leeuw1980}%
  \BibitemOpen
  \bibfield  {author} {\bibinfo {author} {\bibfnamefont {S.~W.}\ \bibnamefont
  {de~Leeuw}}, \bibinfo {author} {\bibfnamefont {J.~W.}\ \bibnamefont
  {Perram}},\ and\ \bibinfo {author} {\bibfnamefont {E.~R.}\ \bibnamefont
  {Smith}},\ }\bibfield  {title} {\bibinfo {title} {Simulation of electrostatic
  systems in periodic boundary conditions. {I}. {L}attice sums and dielectric
  constants},\ }\href {https://doi.org/https://doi.org/10.1098/rspa.1980.0135}
  {\bibfield  {journal} {\bibinfo  {journal} {Proc. R. Soc. A}\ }\textbf
  {\bibinfo {volume} {373}},\ \bibinfo {pages} {27 } (\bibinfo {year}
  {1980})}\BibitemShut {NoStop}%
\bibitem [{\citenamefont {Melko}\ and\ \citenamefont
  {Gingras}(2004)}]{Melko2004}%
  \BibitemOpen
  \bibfield  {author} {\bibinfo {author} {\bibfnamefont {R.~G.}\ \bibnamefont
  {Melko}}\ and\ \bibinfo {author} {\bibfnamefont {M.~J.~P.}\ \bibnamefont
  {Gingras}},\ }\bibfield  {title} {\bibinfo {title} {{M}onte {C}arlo studies
  of the dipolar spin ice model},\ }\href
  {https://doi.org/10.1088/0953-8984/16/43/r02} {\bibfield  {journal} {\bibinfo
   {journal} {Journal of Physics: Condensed Matter}\ }\textbf {\bibinfo
  {volume} {16}},\ \bibinfo {pages} {R1277} (\bibinfo {year}
  {2004})}\BibitemShut {NoStop}%
\bibitem [{\citenamefont {Jaubert}(2009)}]{ThesisJaubert}%
  \BibitemOpen
  \bibfield  {author} {\bibinfo {author} {\bibfnamefont {L.~D.~C.}\
  \bibnamefont {Jaubert}},\ }\emph {\bibinfo {title} {Topological constraints
  and defects in spin ice}},\ \href
  {https://theses.hal.science/tel-00462970/document} {Ph.D. thesis},\ \bibinfo
  {school} {Ecole Normale Sup\'erieure de Lyon} (\bibinfo {year}
  {2009})\BibitemShut {NoStop}%
\bibitem [{\citenamefont {Hall\'en}\ \emph {et~al.}(2022)\citenamefont
  {Hall\'en}, \citenamefont {Grigera}, \citenamefont {Tennant}, \citenamefont
  {Castelnovo},\ and\ \citenamefont {Moessner}}]{Hallen2022}%
  \BibitemOpen
  \bibfield  {author} {\bibinfo {author} {\bibfnamefont {J.~N.}\ \bibnamefont
  {Hall\'en}}, \bibinfo {author} {\bibfnamefont {S.~A.}\ \bibnamefont
  {Grigera}}, \bibinfo {author} {\bibfnamefont {D.~A.}\ \bibnamefont
  {Tennant}}, \bibinfo {author} {\bibfnamefont {C.}~\bibnamefont
  {Castelnovo}},\ and\ \bibinfo {author} {\bibfnamefont {R.}~\bibnamefont
  {Moessner}},\ }\bibfield  {title} {\bibinfo {title} {Dynamical fractal and
  anomalous noise in a clean magnetic crystal},\ }\href
  {https://doi.org/10.1126/science.add1644} {\bibfield  {journal} {\bibinfo
  {journal} {Science}\ }\textbf {\bibinfo {volume} {378}},\ \bibinfo {pages}
  {1218} (\bibinfo {year} {2022})}\BibitemShut {NoStop}%
\bibitem [{\citenamefont {Ryzhkin}(2005)}]{Ryzhkin2005}%
  \BibitemOpen
  \bibfield  {author} {\bibinfo {author} {\bibfnamefont {I.~A.}\ \bibnamefont
  {Ryzhkin}},\ }\bibfield  {title} {\bibinfo {title} {Magnetic relaxation in
  rare-earth oxide pyrochlores},\ }\href {https://doi.org/10.1134/1.2103216}
  {\bibfield  {journal} {\bibinfo  {journal} {Journal of Experimental and
  Theoretical Physics}\ }\textbf {\bibinfo {volume} {101}},\ \bibinfo {pages}
  {481 } (\bibinfo {year} {2005})}\BibitemShut {NoStop}%
\bibitem [{\citenamefont {Jaubert}\ and\ \citenamefont
  {Holdsworth}(2009)}]{Jaubert2009}%
  \BibitemOpen
  \bibfield  {author} {\bibinfo {author} {\bibfnamefont {L.~D.~C.}\
  \bibnamefont {Jaubert}}\ and\ \bibinfo {author} {\bibfnamefont {P.~C.~W.}\
  \bibnamefont {Holdsworth}},\ }\bibfield  {title} {\bibinfo {title} {Signature
  of magnetic monopole and dirac string dynamics in spin ice},\ }\href
  {https://doi.org/10.1038/nphys1227} {\bibfield  {journal} {\bibinfo
  {journal} {Nature Physics}\ }\textbf {\bibinfo {volume} {5}},\ \bibinfo
  {pages} {258} (\bibinfo {year} {2009})}\BibitemShut {NoStop}%
\bibitem [{\citenamefont {Glauber}(1963)}]{Glauber1963}%
  \BibitemOpen
  \bibfield  {author} {\bibinfo {author} {\bibfnamefont {R.~J.}\ \bibnamefont
  {Glauber}},\ }\bibfield  {title} {\bibinfo {title} {Time‐dependent
  statistics of the {I}sing model},\ }\href {https://doi.org/10.1063/1.1703954}
  {\bibfield  {journal} {\bibinfo  {journal} {Journal of Mathematical Physics}\
  }\textbf {\bibinfo {volume} {4}},\ \bibinfo {pages} {294} (\bibinfo {year}
  {1963})}\BibitemShut {NoStop}%
\bibitem [{\citenamefont {S\"uzen}(2014)}]{Suzen2014}%
  \BibitemOpen
  \bibfield  {author} {\bibinfo {author} {\bibfnamefont {M.}~\bibnamefont
  {S\"uzen}},\ }\bibfield  {title} {\bibinfo {title} {Effective ergodicity in
  single-spin-flip dynamics},\ }\href
  {https://doi.org/10.1103/PhysRevE.90.032141} {\bibfield  {journal} {\bibinfo
  {journal} {Phys. Rev. E}\ }\textbf {\bibinfo {volume} {90}},\ \bibinfo
  {pages} {032141} (\bibinfo {year} {2014})}\BibitemShut {NoStop}%
\bibitem [{\citenamefont {Binder}\ and\ \citenamefont
  {Heermann}(2010)}]{Binder2010}%
  \BibitemOpen
  \bibfield  {author} {\bibinfo {author} {\bibfnamefont {K.}~\bibnamefont
  {Binder}}\ and\ \bibinfo {author} {\bibfnamefont {D.~W.}\ \bibnamefont
  {Heermann}},\ }\href {https://doi.org/10.1007/978-3-642-03163-2} {\emph
  {\bibinfo {title} {Monte Carlo Simulation in Statistical Physics: An
  Introduction}}},\ \bibinfo {edition} {5th}\ ed.\ (\bibinfo  {publisher}
  {Springer, New York},\ \bibinfo {year} {2010})\BibitemShut {NoStop}%
\bibitem [{\citenamefont {Fukazawa}\ \emph {et~al.}(2002)\citenamefont
  {Fukazawa}, \citenamefont {Melko}, \citenamefont {Higashinaka}, \citenamefont
  {Maeno},\ and\ \citenamefont {Gingras}}]{Fukazawa2002}%
  \BibitemOpen
  \bibfield  {author} {\bibinfo {author} {\bibfnamefont {H.}~\bibnamefont
  {Fukazawa}}, \bibinfo {author} {\bibfnamefont {R.~G.}\ \bibnamefont {Melko}},
  \bibinfo {author} {\bibfnamefont {R.}~\bibnamefont {Higashinaka}}, \bibinfo
  {author} {\bibfnamefont {Y.}~\bibnamefont {Maeno}},\ and\ \bibinfo {author}
  {\bibfnamefont {M.~J.~P.}\ \bibnamefont {Gingras}},\ }\bibfield  {title}
  {\bibinfo {title} {Magnetic anisotropy of the spin-ice compound
  {D}y$_{2}${T}i$_{2}${O}$_{7}$},\ }\href
  {https://doi.org/10.1103/PhysRevB.65.054410} {\bibfield  {journal} {\bibinfo
  {journal} {Phys. Rev. B}\ }\textbf {\bibinfo {volume} {65}},\ \bibinfo
  {pages} {054410} (\bibinfo {year} {2002})}\BibitemShut {NoStop}%
\bibitem [{\citenamefont {Levis}\ and\ \citenamefont
  {Cugliandolo}(2013)}]{Levis2013}%
  \BibitemOpen
  \bibfield  {author} {\bibinfo {author} {\bibfnamefont {D.}~\bibnamefont
  {Levis}}\ and\ \bibinfo {author} {\bibfnamefont {L.~F.}\ \bibnamefont
  {Cugliandolo}},\ }\bibfield  {title} {\bibinfo {title} {Defects dynamics
  following thermal quenches in square spin ice},\ }\href
  {https://doi.org/10.1103/PhysRevB.87.214302} {\bibfield  {journal} {\bibinfo
  {journal} {Phys. Rev. B}\ }\textbf {\bibinfo {volume} {87}},\ \bibinfo
  {pages} {214302} (\bibinfo {year} {2013})}\BibitemShut {NoStop}%
\bibitem [{\citenamefont {Jaubert}\ and\ \citenamefont
  {Holdsworth}(2011)}]{Jaubert2011}%
  \BibitemOpen
  \bibfield  {author} {\bibinfo {author} {\bibfnamefont {L.~D.~C.}\
  \bibnamefont {Jaubert}}\ and\ \bibinfo {author} {\bibfnamefont {P.~C.~W.}\
  \bibnamefont {Holdsworth}},\ }\bibfield  {title} {\bibinfo {title} {Magnetic
  monopole dynamics in spin ice},\ }\href
  {https://doi.org/10.1088/0953-8984/23/16/164222} {\bibfield  {journal}
  {\bibinfo  {journal} {Journal of Physics: Condensed Matter}\ }\textbf
  {\bibinfo {volume} {23}},\ \bibinfo {pages} {164222} (\bibinfo {year}
  {2011})}\BibitemShut {NoStop}%
\bibitem [{\citenamefont {Stauffer}\ and\ \citenamefont
  {Aharony}(1992)}]{Stauffer2010}%
  \BibitemOpen
  \bibfield  {author} {\bibinfo {author} {\bibfnamefont {D.}~\bibnamefont
  {Stauffer}}\ and\ \bibinfo {author} {\bibfnamefont {A.}~\bibnamefont
  {Aharony}},\ }\href {https://doi.org/https://doi.org/10.1201/9781315274386}
  {\emph {\bibinfo {title} {Introduction To Percolation Theory}}},\ \bibinfo
  {edition} {2nd}\ ed.\ (\bibinfo  {publisher} {Taylor \& Francis, London},\
  \bibinfo {year} {1992})\BibitemShut {NoStop}%
\bibitem [{\citenamefont {Lorenz}\ and\ \citenamefont
  {Ziff}(1998)}]{Lorenz1998}%
  \BibitemOpen
  \bibfield  {author} {\bibinfo {author} {\bibfnamefont {C.~D.}\ \bibnamefont
  {Lorenz}}\ and\ \bibinfo {author} {\bibfnamefont {R.~M.}\ \bibnamefont
  {Ziff}},\ }\bibfield  {title} {\bibinfo {title} {Precise determination of the
  bond percolation thresholds and finite-size scaling corrections for the sc,
  fcc, and bcc lattices},\ }\href {https://doi.org/10.1103/PhysRevE.57.230}
  {\bibfield  {journal} {\bibinfo  {journal} {Phys. Rev. E}\ }\textbf {\bibinfo
  {volume} {57}},\ \bibinfo {pages} {230} (\bibinfo {year} {1998})}\BibitemShut
  {NoStop}%
\bibitem [{\citenamefont {Scullard}\ and\ \citenamefont
  {Jacobsen}(2020)}]{Scullard2020}%
  \BibitemOpen
  \bibfield  {author} {\bibinfo {author} {\bibfnamefont {C.~R.}\ \bibnamefont
  {Scullard}}\ and\ \bibinfo {author} {\bibfnamefont {J.~L.}\ \bibnamefont
  {Jacobsen}},\ }\bibfield  {title} {\bibinfo {title} {Bond percolation
  thresholds on {A}rchimedean lattices from critical polynomial roots},\ }\href
  {https://doi.org/10.1103/PhysRevResearch.2.012050} {\bibfield  {journal}
  {\bibinfo  {journal} {Phys. Rev. Res.}\ }\textbf {\bibinfo {volume} {2}},\
  \bibinfo {pages} {012050} (\bibinfo {year} {2020})}\BibitemShut {NoStop}%
\bibitem [{\citenamefont {Martins}\ and\ \citenamefont
  {Plascak}(2003)}]{Martins2003}%
  \BibitemOpen
  \bibfield  {author} {\bibinfo {author} {\bibfnamefont {P.~H.~L.}\
  \bibnamefont {Martins}}\ and\ \bibinfo {author} {\bibfnamefont {J.~A.}\
  \bibnamefont {Plascak}},\ }\bibfield  {title} {\bibinfo {title} {Percolation
  on two- and three-dimensional lattices},\ }\href
  {https://doi.org/10.1103/PhysRevE.67.046119} {\bibfield  {journal} {\bibinfo
  {journal} {Phys. Rev. E}\ }\textbf {\bibinfo {volume} {67}},\ \bibinfo
  {pages} {046119} (\bibinfo {year} {2003})}\BibitemShut {NoStop}%
\bibitem [{\citenamefont {Gaunt}\ and\ \citenamefont
  {Ruskin}(1978)}]{Gaunt1978}%
  \BibitemOpen
  \bibfield  {author} {\bibinfo {author} {\bibfnamefont {D.~S.}\ \bibnamefont
  {Gaunt}}\ and\ \bibinfo {author} {\bibfnamefont {H.}~\bibnamefont {Ruskin}},\
  }\bibfield  {title} {\bibinfo {title} {Bond percolation processes in d
  dimensions},\ }\href {https://doi.org/10.1088/0305-4470/11/7/025} {\bibfield
  {journal} {\bibinfo  {journal} {Journal of Physics A: Mathematical and
  General}\ }\textbf {\bibinfo {volume} {11}},\ \bibinfo {pages} {1369}
  (\bibinfo {year} {1978})}\BibitemShut {NoStop}%
\bibitem [{\citenamefont {Galam}\ and\ \citenamefont
  {Mauger}(1996)}]{Galam1996}%
  \BibitemOpen
  \bibfield  {author} {\bibinfo {author} {\bibfnamefont {S.}~\bibnamefont
  {Galam}}\ and\ \bibinfo {author} {\bibfnamefont {A.}~\bibnamefont {Mauger}},\
  }\bibfield  {title} {\bibinfo {title} {Universal formulas for percolation
  thresholds},\ }\href {https://doi.org/10.1103/PhysRevE.53.2177} {\bibfield
  {journal} {\bibinfo  {journal} {Phys. Rev. E}\ }\textbf {\bibinfo {volume}
  {53}},\ \bibinfo {pages} {2177} (\bibinfo {year} {1996})}\BibitemShut
  {NoStop}%
\bibitem [{\citenamefont {Jackson}\ \emph {et~al.}(2014)\citenamefont
  {Jackson}, \citenamefont {Lhotel}, \citenamefont {Giblin}, \citenamefont
  {Bramwell}, \citenamefont {Prabhakaran}, \citenamefont {Matsuhira},
  \citenamefont {Hiroi}, \citenamefont {Yu},\ and\ \citenamefont
  {Paulsen}}]{Jackson2014}%
  \BibitemOpen
  \bibfield  {author} {\bibinfo {author} {\bibfnamefont {M.~J.}\ \bibnamefont
  {Jackson}}, \bibinfo {author} {\bibfnamefont {E.}~\bibnamefont {Lhotel}},
  \bibinfo {author} {\bibfnamefont {S.~R.}\ \bibnamefont {Giblin}}, \bibinfo
  {author} {\bibfnamefont {S.~T.}\ \bibnamefont {Bramwell}}, \bibinfo {author}
  {\bibfnamefont {D.}~\bibnamefont {Prabhakaran}}, \bibinfo {author}
  {\bibfnamefont {K.}~\bibnamefont {Matsuhira}}, \bibinfo {author}
  {\bibfnamefont {Z.}~\bibnamefont {Hiroi}}, \bibinfo {author} {\bibfnamefont
  {Q.}~\bibnamefont {Yu}},\ and\ \bibinfo {author} {\bibfnamefont
  {C.}~\bibnamefont {Paulsen}},\ }\bibfield  {title} {\bibinfo {title} {Dynamic
  behavior of magnetic avalanches in the spin-ice compound
  {${\mathrm{Dy}}_{2}{\mathrm{Ti}}_{2}{\mathrm{O}}_{7}$}},\ }\href
  {https://doi.org/10.1103/PhysRevB.90.064427} {\bibfield  {journal} {\bibinfo
  {journal} {Phys. Rev. B}\ }\textbf {\bibinfo {volume} {90}},\ \bibinfo
  {pages} {064427} (\bibinfo {year} {2014})}\BibitemShut {NoStop}%
\bibitem [{\citenamefont {Tomasello}\ \emph {et~al.}(2019)\citenamefont
  {Tomasello}, \citenamefont {Castelnovo}, \citenamefont {Moessner},\ and\
  \citenamefont {Quintanilla}}]{Tomasello2019}%
  \BibitemOpen
  \bibfield  {author} {\bibinfo {author} {\bibfnamefont {B.}~\bibnamefont
  {Tomasello}}, \bibinfo {author} {\bibfnamefont {C.}~\bibnamefont
  {Castelnovo}}, \bibinfo {author} {\bibfnamefont {R.}~\bibnamefont
  {Moessner}},\ and\ \bibinfo {author} {\bibfnamefont {J.}~\bibnamefont
  {Quintanilla}},\ }\bibfield  {title} {\bibinfo {title} {Correlated quantum
  tunneling of monopoles in spin ice},\ }\href
  {https://doi.org/10.1103/PhysRevLett.123.067204} {\bibfield  {journal}
  {\bibinfo  {journal} {Phys. Rev. Lett.}\ }\textbf {\bibinfo {volume} {123}},\
  \bibinfo {pages} {067204} (\bibinfo {year} {2019})}\BibitemShut {NoStop}%
\bibitem [{NIS()}]{NIST:DLMF:10}%
  \BibitemOpen
  \href {https://dlmf.nist.gov/10} {\bibinfo {title} {{\it NIST Digital Library
  of Mathematical Functions, {S}ection 10}}},\ \bibinfo {howpublished} {Release
  1.1.12 of 2023-12-15},\ \bibinfo {note} {{F}.~W.~J. Olver, A.~B. {Olde
  Daalhuis}, D.~W. Lozier, B.~I. Schneider, R.~F. Boisvert, C.~W. Clark, B.~R.
  Miller, B.~V. Saunders, H.~S. Cohl, and M.~A. McClain, eds.}\BibitemShut
  {Stop}%
\end{thebibliography}%
        
\end{document}